\PassOptionsToPackage{nomarkers}{endfloat}
\documentclass[useAMS, usenatbib, referee]{biom}
\setcitestyle{numbers,square,sort&compress}
\usepackage{amsmath,amssymb}
\usepackage{graphicx}
\usepackage{booktabs}
\usepackage{subcaption}
\usepackage{multirow}
\usepackage{hyperref}
\def\bSig\mathbf{\Sigma}
\def\R\mathbb{R}

\title[VAE based approach for parameter estimation in NLME-ODEs]{Variational autoencoder for inference of nonlinear mixed effect models based on ordinary differential equations}

\author
{Zhe Li, Mélanie Prague, Rodolphe Thiébaut, Quentin Clairon\emailx{zhe.li@u-bordeaux.fr, melanie.prague@inria.fr, \\
rodolphe.thiebaut@u-bordeaux.fr, quentin.clairon@u-bordeaux.fr} \\
Univ. Bordeaux, INSERM BPH, U1219, Inria SISTM team, VRI, France}

\begin{document}


\date{}



\pagerange{\pageref{firstpage}--\pageref{lastpage}} 
\volume{XX}
\pubyear{XXXX}
\artmonth{Feburary}


\doi{10.13140/RG.2.2.32869.46567}


\label{firstpage}


\begin{abstract}
We propose a variationl autoencoder approach for identifiability-awared parameter estimation 
in nonlinear mixed-effects models 
based on ordinary differential equations (NLME-ODEs) using longitudinal data 
from multiple subjects. In moderate dimensions, likelihood-based inference via the stochastic approximation EM algorithm (SAEM) 
is widely used, but it relies on Markov Chain Monte-Carlo (MCMC) to approximate subject-specific posteriors. 
As model complexity increases or observations per subject are sparse and irregular, 
performance often deteriorates due to a complex, multimodal likelihood surface which may lead to MCMC convergence difficulties. 
We instead estimate parameters by maximizing the evidence lower bound (ELBO), a regularized surrogate for the marginal likelihood. 
A shared encoder amortizes inference of subject-specific random effects by avoiding per-subject optimization and the use of MCMC. 
Beyond pointwise estimation, we quantify parameter uncertainty using observed-information-based variance estimator 
and verify that practical identifiability of the model parameters is not compromised by nuisance parameters introduced in the encoder. 
We evaluated the method in three simulation case studies (pharmacokinetics, humoral response to vaccination, and TGF-$\beta$ activation dynamics in asthmatic airways) 
and on a real-world antibody kinetics dataset, comparing against SAEM baselines.
\end{abstract}

%



\maketitle


%

\section{Introduction}
\label{s:intro}

Ordinary differential equation (ODE) models are widely used in systems biology,
mathematical medicine, virology, vaccinology,
and pharmacometrics due to their interpretability and predictive power.
However, classical ODEs typically describe an average trajectory and do not capture inter-individual variability in
longitudinal population data, which are often sparse, irregularly sampled, and noisy.
To address this, ODEs are commonly embedded in nonlinear mixed-effects (NLME) models, yielding NLME-ODEs
\citep{Lavielle2014}, 
where subject-specific random effects capture heterogeneity while preserving a shared mechanistic structure.

More formally, NLME-ODEs consider a population of $n$ subjects. 
For subject $i\in\{1,\dots,n\}$, observations are collected at (possibly irregular) times
$t_{ij}$, $j=1,\dots,n_i$, where $n_i$ denotes the number of measurements for subject $i$;
we write $Y_{ij}$ for the measurement taken at time $t_{ij}$.
The dynamics of the subject \( i \) is described by a \( D_X \)-dimensional ODEs:
\[
\begin{cases}
\dfrac{dX_i(t)}{dt} = f_{\theta_i}\bigl(t, X_i(t)\bigr),\\
X_i(0) = x_{i,0}.
\end{cases}
\]
where \( f_{\theta_i} \) is a \( D_X \)-dimensional vector field defining the system dynamics, 
\( \theta_i \) is a vector of subject-specific parameters, and \( x_{i,0} \) denotes the initial condition. 
Inter-individual variability is modeled through a mixed-effects formulation
$
\theta_i = g(\bar{\theta}, b_i),
$
where $\bar{\theta}$ represents the population-level mean and $b_i$ denotes the
subject-specific deviation around this mean, also known as the random effects. Here, $g(\cdot)$ is a (possibly nonlinear) 
transformation function, for example through additive or log-normal parameterizations to enforce constraints such as positivity.
We assume \(b_i \sim p_\phi(b_i) = \mathcal{N}(0,\Omega)\), where \(b_i=(b_{i1},\ldots,b_{i d_b})^\top \in \mathbb{R}^{D_b}\) 
is the \(D_b\)-dimensional vector of random effects and
\(\Omega \in \mathbb{R}^{D_b\times D_b}\) is a symmetric positive-definite covariance matrix.
To guarantee positive definiteness, we parameterize \(\Omega\) via its Cholesky factor
\(\Omega = L_\Omega L_\Omega^\top\).

Our goal is to estimate the population parameters
$
\phi = (\bar{\theta},\, L_\Omega,\, L_\Sigma)
$ 
of dimension $D_{\phi}$
from the noisy and discrete observations $Y_{ij}$ generated according to the following
observation model:
\begin{equation}
\label{eq:observation_model}
Y_{ij} = h_{\theta_i}(X_i(t_{ij})) + \epsilon_{ij}, 
\qquad 
\epsilon_{ij} \sim \mathcal{N}(\mathbf{0},\, \Sigma).
\end{equation}
Here, \( h_{\theta_i} \) is the observation function mapping the $i$-th subject state represented by the ODEs solution \( X_i \) 
to the measure $Y_{ij}$ and 
\(\Sigma \in \mathbb{R}^{D_{n_i} \times D_{n_i}}\) is the covariance matrix of the measurement noise which is parameterized by its Cholesky
factor \(\Sigma = L_\Sigma L_\Sigma^\top\).

As in classical inference problems, NLME-ODEs parameters can be estimated by maximum likelihood, marginalizing over the
random effects $b_i$ viewed as latent variables. However, maximizing the marginal likelihood requires repeated
integration over $b_i$, which is generally intractable in nonlinear ODE models.
Early approaches such as first-order (FO) and first-order conditional estimation (FOCE) rely on local Taylor/Laplace
approximations to obtain a tractable likelihood \citep{Pinheiro1995}. These approximations can be inaccurate under strong nonlinearities,
non-Gaussian random effects, or sparse individual data, settings common in NLME-ODEs; adaptive Gaussian quadrature can
improve accuracy \citep{Prague2013} but becomes computationally prohibitive as the number of
random effects grows.
A widely used method is the stochastic approximation EM (SAEM) algorithm
\citep{delyon1999convergence,Lavielle2007}, which avoids explicit
numerical integration by sampling random effects from their conditional distribution (typically via MCMC) and updating
parameters through stochastic approximation. SAEM is often more accurate and stable than FO/FOCE or quadrature for
nonlinear models and is implemented in Monolix and Pumas \citep{monolix2024r1,pumas2025}. Nonetheless, for sparse data and/or
high-dimensional random effects, SAEM may become computationally expensive and less reliable, motivating scalable
variational alternatives.

SAEM inherits several limitations from its reliance on MCMC in the E-step. In sparse settings, the conditional
posterior of the random effects is weakly informed by the data, which can lead to slow mixing and poor exploration.
As the dimension of the random effects increases, MCMC becomes increasingly expensive and more sensitive to
multimodality, potentially converging to local modes and yielding non-identifiable solutions. Bayesian MCMC-based
inference or MAP estimation faces similar computational and convergence issues in the same regimes
\citep{Lunn2000,Stan2017}.

More fundamentally, NLME-ODEs likelihoods are often highly non-convex because ODE solutions depend nonlinearly on the
parameters, producing rugged objective landscapes and compounding identifiability and optimization difficulties.
These challenges motivate optimization-based alternatives that can regularize inference and scale to complex models.
Variational inference (VI) replaces stochastic sampling with a tractable, parametric approximation of the posterior,
optimized by maximizing a lower bound. In particular, variational autoencoders (VAEs) use neural networks to
parameterize this approximation, offering a scalable and flexible framework for NLME-ODEs inference.

This coupling between NLME-ODEs and neural networks have already been investigated for other purposes.
\cite{Qian2021} proposed to integrates expert pharmacological ODEs into Neural ODEs to provide additional
insights to the clinicians in disease-progression modeling.
\cite{Braem2025} introduced low-dimensional neural ODEs implemented directly in Monolix \citep{monolix2024r1}, 
showing how inter-individual variability 
can be incorporated into neural ODEs parameters while preserving pharmacometric interpretability. 
\cite{Janssen2024} compared FO/FOCE 
and VI approaches for mixed-effect estimation in Deep Compartment Models.
\cite{Arruda2024} developed an amortized inference framework 
in which a conditional normalizing flow is trained on simulated data generated from the parametric ODE model 
to approximate the parameter posterior distribution.
\cite{Martensen2024} introduced a Deep Nonlinear Mixed-Effect \citep{deeppumas2025} modeling framework 
in which a neural network term is embedded directly into the ODE 
to represent unknown system dynamics. 
\cite{roeder2019vihds} performed amortised Bayesian inference for NLME-ODEs
through a conditional VAE with population-, group-, and individual-level latent blocks.

Despite these advances, most deep or variational extensions of NLME-ODEs emphasize predictive performance and scalability
over statistical interpretability and identifiability. Amortized approaches
\citep{roeder2019vihds,Arruda2024} provide efficient posterior approximations but do not explicitly address
identifiability. Conversely, methods that embed neural networks
into pharmacometric ODEs \citep{Braem2025,Martensen2024} typically rely on SAEM or FOCE, and thus remain
computationally demanding and sensitive to initialization; moreover, neural components is generally nonidentifiable due
to redundancy.
Recent ELBO-based inference methods for NLME–ODE models close to 
ours have been proposed by \cite{Rohleff2025VAEOneRun}. However, they do not address practical identifiability and uncertainty quantification, as they do not construct a variance estimator. We focus on these aspects because they are essential for interpretability. 
Moreover, they do not provide a rigorous comparison with SAEM-based methods on these points.

In this paper, we propose a interpretable, identifiability-awared amortized variational inference framework for NLME--ODEs tailored to sparse and irregular
longitudinal settings. We maximize the evidence
lower bound (ELBO), yielding a tractable objective that avoids the sampling step of SAEM. For each subject, the
posterior distribution of random effects is approximated within a parametric family parameterized by a lightweight,
shallow neural encoder to prevent overfitting and non-identifiability given the
limited data; the decoder is a parametric ODE model which reconstructs the trajectory from
the random effects, ensuring that the inferred representations retain a mechanistic
interpretation as subject-specific biological parameters. Although the method is compatible with higher-dimensional
random-effects structures, in this work we primarily evaluate its performance in data-scarce regimes 
and leave a systematic study of scalability with respect to the random-effect dimension
to future work.

Beyond point estimation, we quantify population-level parameter uncertainty via a variance estimator based on the observed
Fisher information matrix, computed by automatic differentiation at the variational optimum for population parameters.
 
Identifiability is fundamental for interpretable inference in NLME models. When neural network components are introduced, 
this property can be compromised by additional nuisance parameters or redundant representations. We further assess convergence behavior 
to verify the practical identifiability of the underlying ODEs, ensuring that each population parameter can be uniquely determined from the data. 
These steps guarantee that the resulting inference remains stable and interpretable within the NLME-ODEs framework.

The paper is organized as follows:
\textbf{Section 2} introduces the proposed method, including the variational inference strategy, 
  identifiability analysis, and uncertainty quantification procedure;
\textbf{Section 3} reports simulation results in which we compare our method with the approach SAEM 
implemented in Monolix \citep{monolix2024r1} in terms of estimation accuracy and identifiability issues under various scenarios, 
in particular for the estimation of NLME-ODEs of increasing complexities in terms of numbers of estimated parameters;
\textbf{Section 4} presents an application to real clinical data on antibody concentration dynamics 
  following immunization with a COVID-19 vaccine as in \citep{Clairon2023}, demonstrating the practical relevance 
  and robustness of the proposed framework.

\section{Amortized Variational Inference for NLME--ODEs}\label{s:model}

\subsection{Marginal Likelihood and Variational Reformulation}
\label{sec:var_reformulation}
Classical inference for NLME models maximizes the population marginal log-likelihood
\begin{equation}
\ell(\phi;\mathbf Y)=\sum_{i=1}^n \log p_\phi(Y_i),
\qquad
p_\phi(Y_i)=\int p_\phi(Y_i\mid b_i)\,p_\phi(b_i)\,db_i,
\end{equation}
where $\mathbf Y=\{Y_i\}_{i=1}^n$ and $Y_i=\{Y_{ij}\}_{j=1}^{n_i}$ are observations at times
$\{t_{ij}\}_{j=1}^{n_i}$. Under the observation model \ref{eq:observation_model}:
\begin{equation}
p_\phi(Y_i\mid b_i)=\prod_{j=1}^{n_i}\mathcal N\!\Big(Y_{ij}-h_{\theta_i}\big(X_i(t_{ij})\big),\,\Sigma\Big),
\end{equation}
where $\mathcal N(\cdot,\Sigma)$ denotes the density of a centered Gaussian with covariance $\Sigma$ evaluated at the residual
$Y_{ij}-h_{\theta_i}(X_i(t_{ij}))$. The integral over $b_i$ is generally intractable, so we adopt variational inference and
introduce
$
q_\psi(b_i\mid Y_i)=\mathcal N\!\Big(\mu_\psi(Y_i),\,L_\psi(Y_i)L_\psi(Y_i)^\top\Big),
$
where \(L_\psi(Y_i)\) is a lower-triangular Cholesky factor (with positive diagonal). This approximation to $p_\phi(b_i\mid Y_i)$ yields the individual ELBO following \citep{kingma2014vae}:
\begin{equation}
\label{eq:ELBO_i}
\mathcal L_{\mathrm{elbo}_i}(\phi,\psi;Y_i)
=\mathbb E_{q_\psi(b_i\mid Y_i)}\!\big[\log p_\phi(Y_i\mid b_i)\big]
-\mathrm{KL}\!\big(q_\psi(b_i\mid Y_i)\,\Vert\,p_\phi(b_i)\big),
\end{equation}
and the population objective is
\begin{equation}
\mathcal L_{\mathrm{ELBO}}(\phi,\psi;\mathbf Y)=\sum_{i=1}^n \mathcal L_{\mathrm{elbo}_i}(\phi,\psi;Y_i)
\le \ell(\phi;\mathbf Y),
\end{equation}
where the inequality follows from Jensen's inequality (see Supplementary Web Materials Section 1 for details).

\noindent In our model, we have two distinct sets of parameters:
\begin{enumerate}
\item structural parameters of interest:  $\phi = (\bar{\theta}, L_{\Omega}, L_{\Sigma}) \in\mathbb R^{D_\phi}$.
\item nuisance parameters: the variational parameter $\psi\in\mathbb R^{D_\psi}$ that parameterizes the approximate posterior $q_\psi$.
\end{enumerate}
So the estimator can be defined as:
\begin{equation}
\label{eq:estimator}
    (\hat{\phi}, \hat{\psi}) = 
    \operatorname*{arg\,max}_{(\phi, \psi)}
    \mathcal{L}_{\mathrm{ELBO}}(\phi, \psi; \mathbf{Y})
\end{equation}

\subsection{Variational Autoencoder}
\label{sec:encoder_design}
To optimize the ELBO, we adopt amortized variational inference as in \citep{kingma2014vae}, where a shared neural network encoder parameterized by $\psi$ 
maps each subject’s longitudinal observations to the parameters of the approximate posterior. 
The expressiveness of the variational family can be adapted to data availability and model complexity.
In this work we use a Gaussian family for computational efficiency, but richer distributions 
(e.g., Gaussian mixtures or normalizing flows) can be employed when greater flexibility is required.
The encoder outputs \(\mu_\psi(Y_i)\) and \(L_\psi(Y_i)\) of the approximate posterior distribution. 
Coupled with an ODEs-based decoder that reconstructs the data from the random effects, 
this architecture forms a VAE adapted to NLME-ODEs. 

Amortized inference offers several advantages in the context of longitudinal data
with sparse available measurement per subject, such as in clinical trial setting. By
sharing statistical strength across subjects, it improves inference stability even when
individual trajectories are short, sparse, or irregularly sampled. Once trained, the
encoder provides near-instant posterior estimates for new subjects, making the approach computationally efficient and scalable to large cohorts. Furthermore, by
learning a smooth mapping from observed data to latent parameters, the amortized encoder regularizes the inverse problem, reducing overfitting and enhancing
generalization to unseen clinical profiles.

The encoder architecture depends on the sampling design of the longitudinal data. 
For regularly sampled trajectories with fixed sequence length, we use a lightweight encoder 
combining one-dimensional convolutional and projection layers 
with smooth nonlinear activation functions such as the Gaussian Error Linear Unit (GELU). 
Given the regular grid, local temporal dependencies can be captured effectively by convolutions, 
and compared with recurrent architectures, this design uses fewer parameters and is therefore 
less prone to overfitting in our setting, while maintaining numerical stability and 
identifiability of the model parameters.
For irregularly sampled trajectories, we instead use a recurrent encoder, which takes as input both the observed values and the elapsed time 
between successive measurements ($\Delta t_{ij} = t_{ij}-t_{i,j-1}$), together with a normalized absolute time covariate. 
These additional time inputs allow the network to adapt its dynamics to heterogeneous sampling 
intervals, which is crucial when the timing of observations carries information about the 
underlying process. Given the limited size of the data in our application, we use shallow architectures with small 
hidden dimensions smaller than $32$ to limit overfitting and avoid introducing non-identifiability due to 
excessive network capacity.

In real-world applications, longitudinal sequences typically have variable lengths.
To enable efficient batching, we pad each sequence to a common length $T$
(the maximum sequence length in the dataset) and define a binary mask
$m_{it}\in\{0,1\}$ indicating whether time step $t$ for subject $i$ is observed
($m_{it}=1$) or corresponds to padding/missingness ($m_{it}=0$). 
The masked attention pooling is applied by aggregating
variable-length sequences into a fixed-dimensional subject representation throughout the encoder.
In the reconstruction term, the mask ensures
that padded positions does not contribute to the objective \citep{vaswani2017attention}.

Given a latent sample \(b_i\), the decoder reconstructs the corresponding subject-specific 
dynamics by integrating the nonlinear ODEs system $X_i(t) = \mathrm{ODESolve}(f_{\theta_i}, x_{i,0}, t).$
We solve the mechanistic ODE systems using adaptive time-stepping methods, selecting the solver
according to the stiffness properties of the dynamics. For non-stiff or mildly stiff systems, we use
an explicit Runge--Kutta method of order five. For models exhibiting
moderate stiffness, we use a fifth-order diagonally implicit Runge--Kutta scheme, 
which provides improved stability in the presence of fast transient modes.
Relative and absolute error tolerances are set adaptively and specified per model in the corresponding
experimental section. All integrations are performed with the \texttt{Diffrax} \citep{kidger2023diffrax} library.

During the training, the parameters are updated using Adam \citep{kingma2015adam}. 
More details of the optimization, in particular the stopping criterion rules, are in the Supplementary Web Materials Section 4.
\subsubsection{Pitfall of posterior collapse}

A well-known failure mode in VAE is posterior collapse \citep{dai2020usualsuspects}, 
where the encoder learns a distribution too close to the prior which is fixed as $\mathcal{N}(\mathbf{0}, I_{D_b})$ and fails to capture individual-level variability. 
In our case, the risk of posterior collapse is inherently reduced due to two modeling choices:
\begin{itemize}
  \item Learned variance of the prior \( \Omega \): Instead of fixing the prior, we jointly learn its covariance structure, which increases flexibility and prevents the prior from dominating the variational posterior;
  \item ODEs decoder conditioned on \( b_i \): The subject-specific random effects \( b_i \) directly influences the ODEs vector field, ensuring it has a non-negligible effect on the data likelihood.
\end{itemize}

\subsection{Parameters Identifiability Analysis}
Parameter identifiability is a fundamental requirement in mixed-effects and dynamical system modeling, 
as it determines whether model parameters can be uniquely inferred from observed data.
In the context of inverse problems, accurate and interpretable parameter estimation depends critically 
on this property: a model is identifiable if distinct parameter values yield distinct distributions of the data.
Formally, identifiability is defined as
\[
\forall\, \phi_1, \phi_2 \in \Phi, \quad 
p_{\phi_1}(\mathbf{Y}) = p_{\phi_2}(\mathbf{Y}) \ \Rightarrow\  \phi_1 = \phi_2.
\]
This ensures that the mapping from parameters $\phi$ to the data distribution $p_\phi(\mathbf{Y})$ is injective.

We distinguish structural identifiability, which concerns the uniqueness of parameters given perfect, 
noise-free observations of the system’s states \citep{Bellman1970, Audoly2001}, 
from practical identifiability, 
which reflects the ability to recover parameters from finite and noisy data \citep{Raue2009,Lavielle2016IdentifiabilityMEM}. 

In this work, we assume that the structural identifiability of the ODEs is established, that is, 
the underlying biological dynamics are parameterized in a way that allows unique recovery of the true parameters under ideal conditions. 
We assess practical identifiability empirically by repeatedly reinitializing 
the estimation procedure from different random parameter values, also called convergence assessement.
Consistent convergence of the learned parameters across runs suggests practical identifiability 
and indicates that the inference algorithm avoids the local optima 
that often arise in classical likelihood-based methods and signal non-identifiability.

\subsection{Uncertainty Quantification of Population-Level Parameters}
\label{sec:variance_estimator}
By assuming that the approximate variational posterior $q_{\hat{\psi}} (b_i|Y_i)$ provides a sufficiently accurate approximation of the true posterior $p_{\phi}$,
we estimate the variance of \(\hat\phi\) using the inverse observed Fisher information derived from the marginal likelihood. 

Following \cite{MargossianBlei2024}, the individual ELBO error decomposes as:
\[
\log p_\phi(Y_i)-\mathcal L_{\mathrm{elbo},i}(\phi,\psi;Y_i)
=
\underbrace{\big[\log p_\phi(Y_i)-\mathcal L(\phi,q^*;Y_i)\big]}_{\text{variational gap}}
+
\underbrace{\big[\mathcal L(\phi,q^*;Y_i)-\mathcal L_{\mathrm{elbo},i}(\phi,\psi;Y_i)\big]}_{\text{amortization gap}},
\]
where \(q^*(b_i\mid Y_i)\in\mathcal Q\) maximizes the ELBO in the chosen variational family. 
The amortization gap arises because a single encoder \(q_\psi(b_i\mid Y_i)\) amortizes inference across
subjects rather than optimizing a separate \(q_i\) for each subject \(i\). \cite{MargossianBlei2024} shows that this
gap can vanish when an ideal inference function exists and the encoder class is sufficiently expressive for
exchangeable latent-variable models with
$
p_\phi(b,\mathbf Y)=\prod_{i=1}^n p_\phi(b_i)\,p_\phi(Y_i\mid b_i).
$
Our NLME--ODE model satisfies this factorization since \(b_i\sim\mathcal N(\mathbf 0,\Omega)\) are i.i.d.\
given \(\phi\). Hence an ideal mapping \(f^*(Y_i)\) to the optimal variational parameters exists, and when
\(q_\psi(b_i\mid Y_i)\) approximates \(f^*\) within the Gaussian family, the amortization gap is negligible.

The variational
gap equals \(\mathrm{KL}\!\left(q^*(b_i\mid Y_i)\,\Vert\,p_\phi(b_i\mid Y_i)\right)\) and generally does not
vanish for a fixed \(\mathcal Q\) due to variational misspecification.
In our framework, we do not place a prior distribution on the population
parameters \(\phi\); instead, \((\phi,\psi)\) are learned by maximizing the population-level ELBO:
$
(\hat\phi,\hat\psi)=\arg\max\limits_{\phi,\psi} \frac1n\sum_{i=1}^n \mathcal L_{\mathrm{elbo},i}(\phi,\psi;Y_i),
$
so \((\hat\phi,\hat\psi)\) is an M-estimator with \(\psi\) a shared nuisance parameter.
Equivalently, \(\hat\phi\) maximizes the profiled criterion
$
M_n(\phi) =\frac1n\sum_{i=1}^n \mathcal L_{\mathrm{elbo},i}(\phi,\hat\psi;Y_i).
$
If the variational gap (and its \(\phi\)-derivatives) is uniformly small near the true parameter, maximizing
the ELBO is close to maximizing \(\frac1n\sum_{i=1}^n \log p_\phi(Y_i)\), so that \(\hat\phi\) behaves like a
quasi-MLE.

When both gaps are small, the population ELBO tightly approximates the marginal log-likelihood. 
In this regime, we quantify uncertainty in \(\hat\phi\) by approximating \(\log p_\phi(Y_i)\) via Monte Carlo over
\(b_i\) and computing the Hessian at \(\hat\phi\), yielding the observed information
\begin{equation}
I_n(\hat\phi)=-\sum_{i=1}^n \nabla_\phi^2 \log p_{\hat\phi}(Y_i),
\label{eq:observed_FIM}
\end{equation}
where \(p_{\hat\phi}(Y_i)\) denotes the marginal likelihood for subject \(i\) at \(\hat\phi\).

To evaluate $p_{\hat{\phi}}(Y_i)$ and its derivatives, we apply a Monte Carlo approximation
based on the reparameterization trick.
We express the random effects $b_i$ as a transformation
of a noise variable $p_{\epsilon} \sim \mathcal{N}(\mathbf{0}, I_{D_b})$ independent of $\phi$:
$b_i = \mathcal{T}_{\hat{\phi}}(\epsilon) = L_{\hat{\Omega}}\epsilon$
where $L_{\hat{\Omega}}$ is the estimated lower-triangular Cholesky factor  
of the prior distribution $p_{\hat{\phi}}(b_i)$. So we can
approximate the marginal likelihood using monte carlo as:
\begin{equation}
p_{\hat{\phi}}(Y_i) \approx 
\frac{1}{L}\sum_{\ell=1}^{L} p_{\hat{\phi}}\bigl(Y_{i}\mid b_i^{(l)} \bigr) 
= \frac{1}{L}\sum_{\ell=1}^{L} p_{\hat{\phi}}\bigl(Y_{i}\mid \mathcal{T}_{\hat{\phi}}(\epsilon^{(l)}) \bigr),
\qquad p_{\epsilon^{(l)}} \sim \mathcal{N}(\mathbf{0}, I_{D_b}).
\end{equation}
The corresponding gradient and Hessian of the marginal likelihood are:
\[
\nabla_{\phi} p_{\hat{\phi}}(Y_i)
 = \mathbb{E}_{p(\epsilon)}\big[\nabla_{\phi} p_{\hat{\phi}}(Y_i \mid \mathcal{T}_{\hat{\phi}}(\epsilon))\big]
  \approx \frac{1}{L}\sum_{\ell=1}^{L} \nabla_{\phi} p_{\hat{\phi}}(Y_i \mid \mathcal{T}_{\hat{\phi}}(\epsilon^{(l)})),\\[3pt]
 \]
 \[ 
\nabla_{\phi}^2 p_{\hat{\phi}}(Y_i)
 = \mathbb{E}_{p(\epsilon)}\big[\nabla_{\phi}^2 p_{\hat{\phi}}(Y_i \mid \mathcal{T}_{\hat{\phi}}(\epsilon))\big]
  \approx \frac{1}{L}\sum_{\ell=1}^{L} \nabla_{\hat{\phi}}^2 p_{\hat{\phi}}(Y_i \mid \mathcal{T}_{\hat{\phi}}(\epsilon^{(l)})).
\]
Finally, 
The Hessian of the log-marginal likelihood is computed as:
\begin{equation}
\nabla^2_\phi \log p_{\hat{\phi}}(Y_i) = \frac{\nabla^2_\phi p_{\hat{\phi}}(Y_i) \cdot p_{\hat{\phi}}(Y_i) - \nabla_\phi p_{\hat{\phi}}(Y_i) \nabla_\phi p_{\hat{\phi}}(Y_i)^\top}{p_{\hat{\phi}}(Y_i)^2}.
\end{equation}
\subsection{Subject-specific parameter estimators}
In NLME models, after estimating the population parameters, likelihood-based methods provide
subject-specific parameter estimates known as Empirical Bayes estimates,
defined as the maximizers of the estimated a posteriori distributions
$
\hat{b}_i^{\mathrm{EBE}}
= \arg\max_{b_i} \, p_{\hat{\phi}}(b_i \mid Y_i)
$.
In our framework, the trained encoder provides the variational approximation
$q_{\hat{\psi}}(b_i\mid Y_i)$ and yields EBEs in a single forward pass.
Since $q_{\hat{\psi}}(b_i\mid Y_i)$ is Gaussian with mean $\mu_{\hat{\psi}}(Y_i)$, we use
$
\hat b_i^{\mathrm{VAE}} := \arg\max_{b_i} q_{\hat{\psi}}(b_i\mid Y_i) = \mu_{\hat{\psi}}(Y_i),
$
which provides scalable subject-specific estimates without per-subject optimization.

\section{Simulations}
\label{sec:simulation}
To validate our approach, we evaluate its performance and compare it with a classic likelihood/SAEM based approach 
on several benchmarcks constituted of synthetic data simulated from NLME-ODEs with known ground-truth parameters.
In the simulation study, we consider the special case of independent random effects, i.e.,
\(\Omega = \mathrm{diag}(\omega_1^2,\ldots,\omega_{d_b}^2)\),
and i.i.d. observation noise with diagonal covariance
\(\Sigma = \mathrm{diag}(\sigma_1^2,\ldots,\sigma_{d_{n_i}}^2)\).
We assess the accuracy of pointwise estimation via a Monte-Carlo approach in which $N_{\text{MC}} = 100$ trials of simulated data 
are constituted from which parameters are estimated. While larger Monte Carlo sample sizes (e.g., 
$N_{\text{MC}} = 1000$) are often recommended, we found 
$N_{\text{MC}} = 100$ to provide stable estimates given the substantial computational cost of each fit.
From the estimated parameters $\hat{\phi}_{\text{VAE}_k}$ 
and $\hat{\phi}_{\text{LL}_k}$ 
where $k=1,...N_{\text{MC}}$ obtained from our approach and Monolix (based on SAEM) respectively, 
we compare pointwise estimation accuracy using three standard metrics:

\begin{itemize}
    \item Relative Root Mean Squared Error:
    $
    \text{RRMSE} = \frac{1}{\phi} \sqrt{ \frac{1}{N_{\text{MC}}} \sum_{k=1}^{N_{\text{MC}}} \left( \hat{\phi}_k - \phi \right)^2 };
    $    
    \item Relative Bias:
    $
    \text{Rel.Bias} = \frac{1}{\phi} \left( \frac{1}{N_{\text{MC}}} \sum_{k=1}^{N_{\text{MC}}} \hat{\phi}_k - \phi \right);
    $

    \item Empirical Variance:
    $
    \text{Emp.Var.} = \frac{1}{N_{\text{MC}}} \sum_{k=1}^{N_{\text{MC}}} \hat{\phi}_k^2 - \left( \frac{1}{N_{\text{MC}}} \sum_{k=1}^{N_{\text{MC}}} \hat{\phi}_k \right)^2.
    $
\end{itemize}

We compute the estimated variance (Est. Var) for our approach described in Section \ref{sec:variance_estimator} 
for each obtained $\hat{\phi}_{\text{VAE}_k}$ and the classic Fisher based one available for $\hat{\phi}_{\text{LL}_k}$. 
The accuracy of the variance estimation is then assessed by comparison of the theoretical variance approximated 
this way with the empirical one described before which are supposed to estimate the same quantity.
We also evaluate the frequentist coverage rate of the 95\% confidence interval 
derived from the Est. Var and Emp. Var (Est. Cov and Emp. Cov). 
We systematically evaluate practical identifiability for each inference procedure. 

All experiments were run on a single machine equipped with an NVIDIA A100 GPU (40 GB), using float64 precision. 
Importantly, the proposed VAE-based NLME-ODE approach is not GPU-intensive: 
we were able to run the full pipeline on a standard laptop (Intel Core i7 CPU with an NVIDIA RTX 2000 GPU) including the variance estimation step, 
though it is slower than on the A100.
For the SAEM benchmark, we used \texttt{Monolix 2023R1} on a high-performance CPU server equipped 
with 2x 16-core Skylake Intel 
Xeon Gold 6142 @ 2.6 GHz and 384 GB of RAM. The implementation details are in the Supplementary Web Materials Section 5.
Due to space constraints, the details and results for the pharmacokinetics simulation study are reported in Supplementary Web Materials Section 5.1.

\subsection{Antibody kinetics}
\label{sec:simu_antibody}
\paragraph{Simulation setting}
We designed a simulation study to assess the performance of our VAE-based inference framework:
a partially observed NLME-ODEs describing antibody kinetics following multiple vaccine injections developed in \cite{Clairon2023}. 
The mechanistic model for the given $i$-th subject is defined as:
\begin{equation}
\begin{cases}
\dot{S}_i(t) = \bar{f}_{M_{k,i}} e^{-\delta_V (t - t_k)} - \delta_S S_i(t), \\
\dot{Ab}_i(t) = \vartheta_i S_i(t) - \delta_{Ab} Ab_i(t), \\
(S_i(0), Ab_i(0)) = (0.01, 0.1) 
\label{eq:antibody}
\end{cases}
\end{equation}
with three injection times ${t}_{k=1,2,3}$. 
The above system represents a two-compartment mechanistic model capturing the post-vaccination antibody response dynamics. The state variable 
$S_i$ denotes the secreting cells population responsible for antibody production, while 
$Ab_i$ denotes blood circulating antibody concentration.
The parameters $\bar{f}_{M_k}$ represents the fold-change of B-cells population magnitude able to differentiate into secreting cells 
after k-th injection
compared to the first one (by definition $\bar{f}_{M_{1,i}} = 1$).
The decay rate $\delta_S$ controls the lifespan of the secreting cell population, and 
$\delta_{Ab}$ governs the antibody degradation rate. $\delta_V$ is the degradation rate of the vaccine induced antigen.
The parameter $\vartheta_i$ represents the subject-specific antibody production rate per secreting cells, 
capturing inter-individual variability in immune response intensity.
The pair $(\delta_S, \delta_{Ab})$ is not structurally identifiable when relying solely on $Ab_i$ observations. 
To break the symmetry that causes structural non-identifiability, we impose the constraint:
$
\text{log}(\delta_{Ab}) = \text{log}(\delta_S + \exp(\lambda)),
$
where $\lambda \in \mathbb{R}$ is a log-gap parameter enforcing $\delta_{Ab} > \delta_S$ in the ODEs Eq. \ref{eq:antibody} (more details in Supplementary Web Materials Section 3).

Inter-individual variability is introduced in the parameters $\vartheta_i$ and $\bar{f}_{M_{2,i}}$ according to
\[
\log(\vartheta_i) = \log(\vartheta) + b_{\vartheta,i}, 
\qquad 
b_{\vartheta,i} \sim \mathcal{N}(0, \omega^2_\vartheta),
\]
\[
\log(\bar{f}_{M_2,i}) = \log(\bar{f}_{M_2}) + b_{\bar{f}_{M_{2,i}}}, 
\qquad 
b_{\bar{f}_{M_2},i} \sim \mathcal{N}(0, \omega^2_{\bar{f}_{M_{2,i}}}), 
\]
where $\vartheta$ and $\bar{f}_{M_2}$ are shared across the population. 
In this system, only the antibody compartment is observed, 
whereas the upstream driver of the dynamics remains unmeasured.
The observed antibody concentration is related to the latent state through a log-transformed observation model with $\sigma_{Ab}$:
$
Y_{Ab,ij} = \log_{10} Ab_{ij}(t_{ij}) + \epsilon_{Ab,ij}, 
$ where
$
\epsilon_{Ab,ij} \sim \mathcal{N}(0, \sigma^2_{Ab}).
$
We generated two sets of datasets: $100$ datasets with $15$ measurements per subject 
collected at equally spaced time points, 
and $100$ datasets with 10 measurements per subject 
collected at irregular time points over the 400-day follow-up as illustrated in Figure \ref{fig:all_data}(a) and (b). 
Each dataset consists of $50$ subjects. The true parameter values are in the Supplementary Web Materials Table 1.
\begin{figure}[!htbp]
  \centering
  \includegraphics[width=1\textwidth]{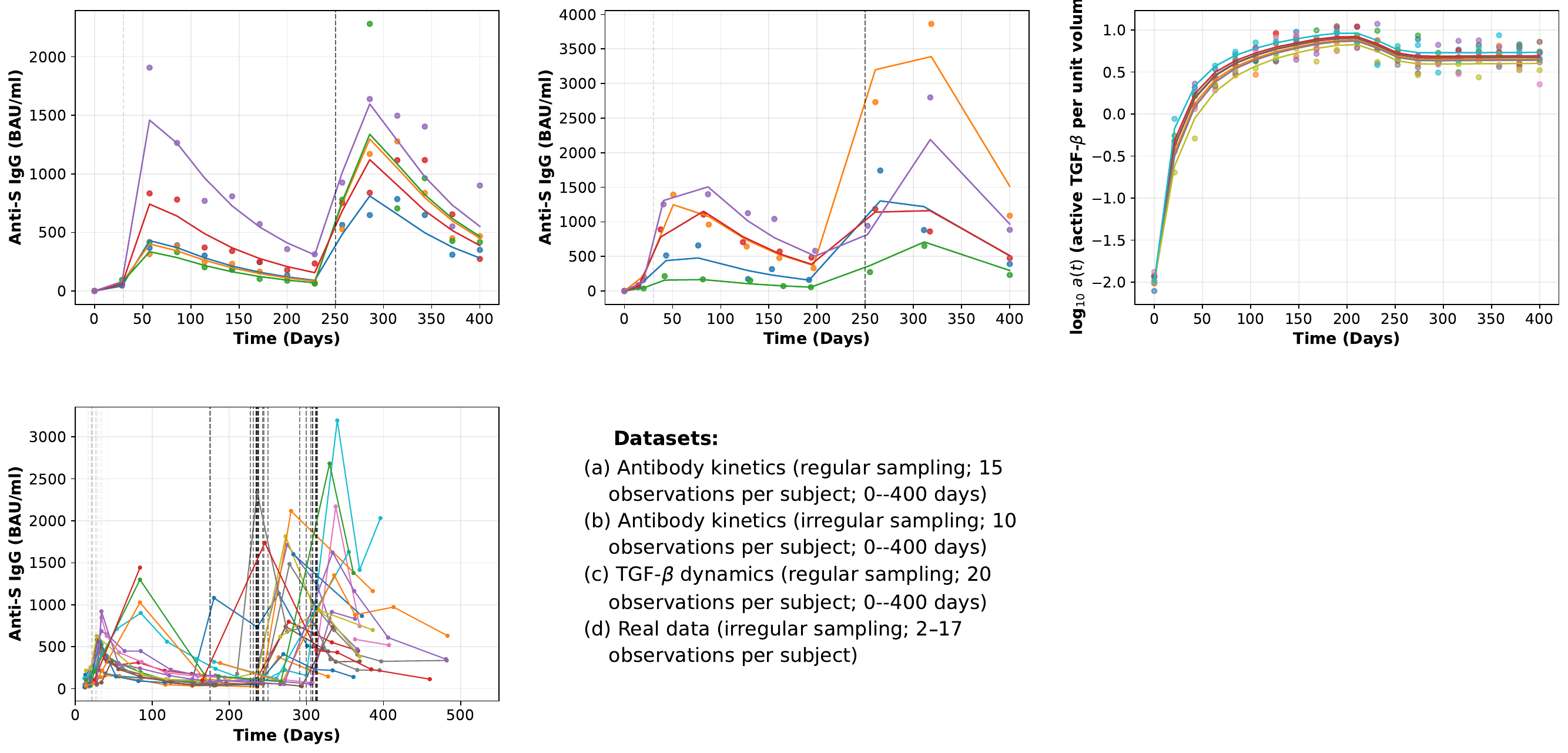}
  \caption{Randomly selected subject trajectories from the simulated datasets and the real dataset. Points are noisy observations; 
  solid curves show the corresponding noise-free underlying dynamics. Vertical light-grey and black dashed lines mark the second and third vaccine injections (the first occurs at 
t=0).}
  \label{fig:all_data}
\end{figure}
\paragraph{Estimation scenario}
The parameters, such as $\delta_{S}$ and $\bar{f}_{M_k}$ affect the observations only indirectly.
Moreover, compensatory effects between \(\delta_{Ab}\) and
\(\delta_S\) can induce strong correlations in their estimates, further impairing identifiability. We therefore consider three inference
scenarios of increasing dimensionality, reflecting progressively more complex settings: 
(S1) baseline scenario in which we estimate the population parameter:
$
\phi = (\log(\vartheta), \log(\bar{f}_{M_2}), \log(\bar{f}_{M_3}), \\\omega_{\vartheta}, \omega_{\bar{f}_{M_2}}, \sigma_{Ab})
$
(S2) scenario S1 augmented with the estimation of the degradation rate of Secreting cells $\log(\delta_S)$
(S3) Scenario S2 is further extended with the antibody decay estimation, $\log(\delta_{Ab})$.  
\paragraph{Results and identifiability analysis}
In the sparse regular sampling setting, the VAE approach yields lower or comparable relative bias, RRMSE and variance of the estimations 
for the population parameters compared to SAEM, confirming its accuracy in recovering fixed effects in the scenario 1. 
However, it performs slightly worse for the random-effect variances, where a modest underestimation is observed. 
This behavior is expected, as variational inference typically produces a tighter posterior approximation 
that tends to underestimate uncertainty components \citep{blei2017variational}.
In Scenario 2, SAEM exhibits a clear inconsistency between its estimated variances and the empirical variances computed across 
the $N_{\text{MC}}=100$ simulated datasets for the mean population parameter $\log(\vartheta), \log(\bar{f}_{M_2})$ and $\log(\bar{f}_{M_3})$. 
The disagreement can be attributed to the FIM computation in SAEM, 
which relies on local curvature at the converged point. In this nonlinear setting with correlated parameters, 
the MCMC-based E-step may mix slowly and drive SAEM toward a suboptimal local basin, 
where the likelihood appears spuriously sharp in certain directions, resulting in variance underestimation.

\begin{figure}[!htbp]
    \centering
    \includegraphics[width=1\textwidth]{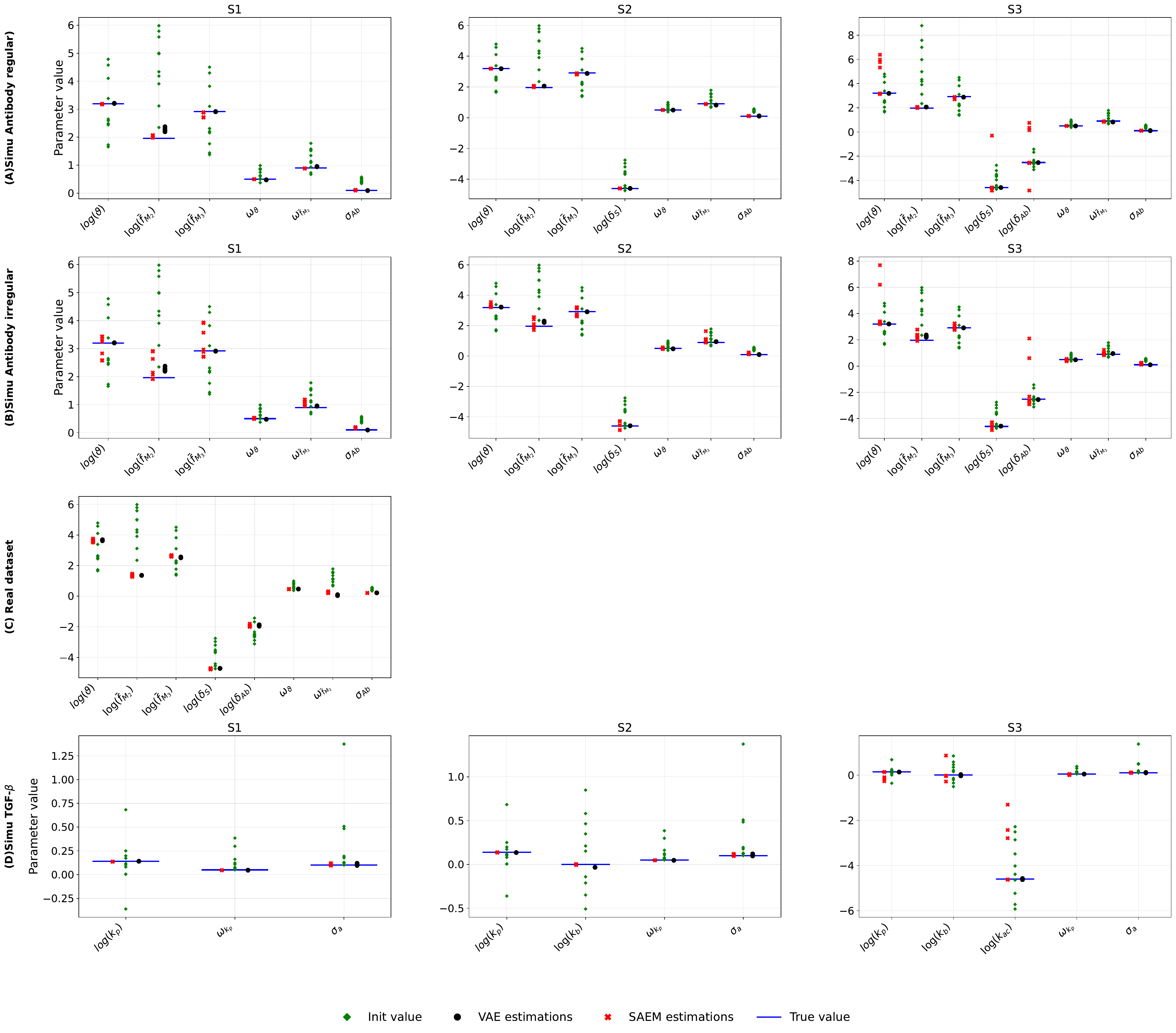}
    \caption{Convergence assessment of SAEM and VAE across three simulation scenarios (S1--S3) and one real-data analysis.
Columns correspond to simulation scenarios (S1, S2, S3). Rows correspond to datasets/cases:
(A) antibody kinetics with regular sampling, (B) antibody kinetics with irregular sampling,
(C) real antibody dataset, and (D) TGF-$\beta$ dynamics.
For each parameter, points show estimates obtained from multiple random initializations for each method.
In simulation panels (A,B,D), the horizontal segment indicates the true parameter value; a tight cluster around it suggests robust convergence,
whereas multiple separated clusters indicate competing local optima and sensitivity to initialization.
In the real-data panel (C), no true value is available, so convergence is assessed by the concentration of the estimates across initializations.}
    \label{fig:Figure_antibody_identifiability}
\end{figure}

In the more complex scenario 3, SAEM frequently converges to local minima and fails to recover several parameters 
as shown in Figure \ref{fig:Figure_antibody_identifiability}(A), 
preventing replication over $N_{\text{MC}}=100$ datasets. 
The VAE method, by contrast, remains numerically stable and produces coherent parameter estimates, 
demonstrating its robustness in moderately ill-conditioned inference regimes.

In the sparse irregular sampling setting, the SAEM algorithm exhibits a loss of practical
identifiability across all three scenarios. This behavior becomes more pronounced as the scenarios become more challenging, and is
particularly visible for parameters governing the decay dynamics
(e.g. $\delta_S$, $\delta_{Ab}$), for which SAEM frequently converges away from the true value.
By contrast, the VAE-based method remains stable under the same sparse irregular design.
We report only the results for Scenario~3 obtained with the VAE in the Table \ref{tab:simulation_results}. 
For SAEM, a non-negligible fraction of simulation runs did not yield reliable standard
error estimates because the observed Hessian was singular for key parameters.
In particular, in Scenario~2 and Scenario~3, the Hessian blocks associated with $\delta_S$ and
$\delta_{Ab}$ were not invertible for approximately $15\%$ and $20\%$ of the datasets respectively,
preventing computation of the estimated covariance matrix and downstream uncertainty summaries.

Overall, these results confirm that partial observation combined with sparse sampling amplifies the practical 
identifiability challenges of NLME-ODEs. 
In such settings, SAEM's reliance on finite-sample MCMC approximations 
in complex posterior spaces can lead to unstable or biased estimation. 
The VAE approach, by directly optimising a smoother ELBO, 
avoids the sampling bottleneck and better preserves identifiability as the parameter space grows.

\begin{table}[!htbp]
\centering
\caption{Comparison of parameter estimates obtained with NN-NLME and SAEM methods on the antibody kinetics simulation datasets in the section \ref{sec:simu_antibody}.
}
\resizebox{\linewidth}{!}{%
\begin{tabular}{cccrrrrrr}
\toprule
\textbf{Scenario} & \textbf{Parameter} & \textbf{Method} &
\multicolumn{1}{c}{\shortstack{\textbf{Rel. Bias}\\($\%$)}} &
\multicolumn{1}{c}{\shortstack{\textbf{RRMSE}\\($\%$)}} &
\multicolumn{1}{c}{\shortstack{\textbf{Emp. Var.}\\($10^{-2}$)}} &
\multicolumn{1}{c}{\shortstack{\textbf{Est. Var.}\\($10^{-2}$)}} &
\textbf{Emp. Cov.} & \textbf{Est. Cov.} \\
\midrule
\multicolumn{9}{c}{\textbf{Antibody kinetics (regular sampling)}}\\
\midrule
S1 & $\log(\vartheta)$      & NN-ELBO & -0.49 & 2.46 & 0.59 & 0.48 & 0.96 & 0.91 \\
    &                     & SAEM & -1.08 & 2.67 & 0.61 & 0.61 & 0.98 & 0.90 \\[4pt]
    & $\log(\bar{f}_{M_2})$     & VAE &  1.38 & 7.12 & 1.87 & 1.29 & 0.94 & 0.90 \\
    &                     & SAEM & 1.65 & 7.22 & 1.90 & 1.81 & 0.95 & 0.93 \\[4pt]
    & $\log(\bar{f}_{M_3})$     & VAE &  0.34 & 1.33 & 0.14 & 0.14 & 0.94 & 0.94 \\
    &                     & SAEM &  1.02 & 1.70 & 0.16 & 0.17 & 0.92 & 0.77 \\[4pt]
    & $\omega_\vartheta$     & VAE & -3.67 & 11.72 & 0.31 & 0.21 & 0.94 & 0.82 \\
    &                     & SAEM & 1.36 & 11.28 & 0.31 & 0.26 & 0.97 & 0.93 \\[4pt]
    & $\omega_{\bar{f}_{M_2}}$  & VAE & -2.41 & 10.58 & 0.86 & 0.68 & 0.94 & 0.86 \\
    &                     & SAEM & 0.11 & 11.23 & 1.02 & 0.88 & 0.94 & 0.88 \\
    &$\sigma_{Ab}$     & VAE & 1.56 & 2.90 & 0.06 & 0.06 & 0.94 & 0.94 \\
    &                     & SAEM & 1.52 & 3.05  & 0.07 & 0.07 & 0.95 & 0.94 \\
\midrule
S2 & $\log(\vartheta)$      & VAE & -0.96 & 2.25 & 0.42 & 0.45 & 0.90 & 0.90 \\
    &                     & SAEM & -0.91 & 2.55 & 0.58 & 1.09 & 0.98 & 0.94 \\[4pt]
    & $\log(\bar{f}_{M_2})$     & VAE &  1.55 & 6.35 & 1.46 & 1.46 & 0.94 & 0.90 \\
    &                     & SAEM &  1.07 & 7.08 & 1.90 & 2.83 & 0.95 & 0.95 \\[4pt]
    & $\log(\bar{f}_{M_3})$     & VAE &  0.60 & 1.39 & 0.13 & 0.14 & 0.95 & 0.92 \\
    &                     & SAEM &  0.78 & 1.59 & 0.16 & 1.15 & 0.90 & 0.83 \\[4pt]
    & $\log(\delta_s)$    & VAE &  0.12 & 0.39 & 0.03 & 0.03 & 0.95 & 0.94 \\
    &                     & SAEM &  0.04 & 0.44 & 0.04 & 0.06 & 0.94 & 0.96 \\[4pt]
    & $\omega_\vartheta$     & VAE & -3.01 & 11.54 & 0.31 & 0.20 & 0.94 & 0.85 \\
    &                     & SAEM & -1.51 & 10.38 & 0.26 & 0.26 & 0.95 & 0.95 \\[4pt]
    & $\omega_{\bar{f}_{M_2}}$  & VAE & -3.19 & 10.98 & 0.89 & 0.85 & 0.92 & 0.88 \\
    &                     & SAEM & $\sim$0.0 & 11.32 & 0.94 & 0.94 & 0.94 & 0.89 \\
    &$\sigma_{Ab}$     & VAE & 1.65 & 3.12 & 0.07 & 0.07 & 0.94 & 0.94 \\
    &                     & SAEM & 1.75 & 3.33  & 0.08 & 0.08 & 0.95 & 0.94 \\
\midrule
S3 & $\log(\vartheta)$      & VAE & -0.70 & 2.42 & 0.55 & 0.73 & 0.93 & 0.93 \\[4pt]
    & $\log(\bar{f}_{M_2})$     & VAE &  0.62 & 6.86 & 1.79 & 1.46 & 0.94 & 0.90 \\[4pt]
    & $\log(\bar{f}_{M_3})$     & VAE &  0.59 & 1.38 & 0.13 & 0.17 & 0.95 & 0.95 \\[4pt]
    & $\log(\delta_s)$    & VAE & 0.14 & 0.52 & 0.05 & 0.06 & 0.93 & 0.95 \\[4pt]
    & $\log(\delta_{Ab})$ & VAE & $\sim$0.0 & 1.97 & 0.25 & 0.24 & 0.95 & 0.95 \\[4pt]
    & $\omega_\vartheta$     & VAE & -2.81 & 11.43 & 0.31 & 0.25 & 0.95 & 0.90 \\[4pt]
    & $\omega_{\bar{f}_{M_2}}$  & VAE & -2.96 & 10.73 & 0.86 & 0.72 & 0.95 & 0.88 \\[4pt]
    &$\sigma_{Ab}$     & VAE & 1.2 & 2.91 & 0.07 & 0.07 & 0.94 & 0.94 \\
\midrule
\multicolumn{9}{c}{\textbf{Antibody kinetics (irregular sampling)}}\\
\midrule
S3 & $\log(\vartheta)$      & VAE & -0.58 & 3.02 & 0.89 & 0.86 & 0.96 & 0.95 \\[4pt]
    & $\log(\bar{f}_{M_2})$     & VAE &  -0.2 & 6.95 & 1.85 & 1.92 & 0.96 & 0.97 \\[4pt]
    & $\log(\bar{f}_{M_3})$     & VAE &  0.14 & 1.63 & 0.22 & 0.29 & 0.95 & 0.96 \\[4pt]
    & $\log(\delta_s)$    & VAE & 0.34 & 0.52 & 0.20 & 0.19 & 0.98 & 0.97 \\[4pt]
    & $\log(\delta_{Ab})$ & VAE & -1.01 & 3.82 & 0.95 & 0.97 & 0.98 & 0.97 \\[4pt]
    & $\omega_\vartheta$     & VAE & -3.65 & 12.97 & 0.39 & 0.34 & 0.90 & 0.87 \\[4pt]
    & $\omega_{\bar{f}_{M_2}}$  & VAE & -4.75 & 14.28 & 1.46 & 1.04 & 0.98 & 0.88 \\[4pt]
    &$\sigma_{Ab}$     & VAE & 1.68 & 3.92 & 2.65 & 0.69 & 0.96 & 0.90 \\
\bottomrule
\end{tabular}
}
\label{tab:simulation_results}
\end{table}


\subsection{Dynamics of TGF Activation in Asthmatic Airways}
\label{sec:asthme_simu}
To further challenge the proposed inference methods, we next consider a more complex ODE system with richer nonlinear interactions.
To reproduce the dynamics of airway remodeling, we adopt the reduced system proposed by
\cite{Pybus2023}, which captures the fast biochemical activation of transforming growth factor~$\beta$ (TGF-$\beta$) 
mediated by airway smooth muscle (ASM) and extracellular matrix (ECM) interactions.  
The model describes the temporal evolution of the densities of proliferating ASM cells $p(t)$,
contractile ASM cells $c(t)$, active TGF-$\beta$ concentration $a(t)$, and ECM $m(t)$ as follows:
\begin{equation}
\begin{cases}
\label{eq:TGF_reduced}
\frac{dp_i}{dt} = 
{\kappa_p}_i p_i\!\left(1 - \frac{p_i}{p_{\max}}\right)
\!\left(1 + \frac{\kappa_{ap} a_i p_i}{\tilde{\eta}_{ap} + a_i p_i}\right)
+ \kappa_{cp}\frac{\gamma_c}{\gamma_p} c_i - \kappa_{pc} p_i, 
\\[4pt]
\frac{dc_i}{dt} = 
\kappa_{pc}\frac{\gamma_p}{\gamma_c} p_i 
- \bigl(\kappa_{cp} + \phi_c\bigr)c_i, 
\\[4pt]
\frac{da_i}{dt} = 
\left(
\frac{\kappa_s}{\nu}
e^{-\frac{(t - \tau_i)^2}{\nu^2}}
+ \frac{\kappa_{ac} a_i c_i}{\tilde{\eta}_{ac} + a_i c_i}
\right)
\frac{c_i} {m_i}{\gamma_a}
- \kappa_b\!\left(\frac{\gamma_p}{\gamma_c}p_i + c_i\right)a_i - a_i, 
\\[4pt]
\frac{dm_i}{dt} = 
\left(
\kappa_{pm} + \frac{\kappa_{apm} a_i p_i}{\tilde{\eta}_{apm} + a_i p_i}
\right)
\gamma_p p_i - \phi_m m_i \\[4pt]
(p_i(0), c_i(0), a_i(0), m_i(0))  = (0.1,\, 0.8,\, 0.01,\, 0.9).
\end{cases}
\end{equation}
The design was implemented with one random effect on the kinetic parameter 
$k_p$ as:
\[
\text{log}(k_{p_i}) = \text{log}(k_p) + b_i \quad \text{and} \quad b_i \sim \mathcal{N}(0, \omega^2_{k_p}), 
\]
representing the individual rate of ASM proliferation. 
Only TGF-beta is observed according to the observation model:
$
 Y_{ij} = \text{log}_{10}(a(t_{ij})) + \epsilon_{a, ij}
$
where $\epsilon_{a, ij} \sim \mathcal{N}(0, \sigma^2_a)$
with 20 longitudinal measurements uniformly distributed over 400 days, example of randomly drawn data 
are represented in Figure \ref{fig:all_data}(c). Each dataset consists of 50 subjects. The true parameters are in the Supplementary Web Materials Table 3.

\paragraph{Estimation scenarios}
The inference task is made difficult by the strongly coupled, feedback-rich ODE structure (logistic proliferation,
phenotype switching, Hill-type saturations, and multiplicative terms such as \(a_i c_i\)), which induces nonconvex
likelihoods, strong parameter interactions, and practical near-nonidentifiabilities. We consider three
inference scenarios of increasing dimensionality: (S1) a baseline setting estimating
\(\phi=\{\log(k_p),\omega_{k_p},\sigma_a\}\); (S2) S1 augmented by estimating \(\log(k_b)\); and (S3) S2 further
extended by estimating \(\log(k_{ac})\).

\paragraph{Results} 
As shown in Table \ref{tab:asthme_results}, the VAE approach consistently outperforms SAEM, 
with smaller relative bias and lower RRMSE across all parameters. 
Regarding practical identifiability (Figure \ref{fig:Figure_antibody_identifiability}D), parameters appear identifiable in Scenarios~1--2, 
but SAEM becomes unstable in Scenario~2: FIM-based coverage for $k_b$ drops to $2\%$ due to severe variance underestimation. 
This likely reflects a challenging nonconvex likelihood, causing SAEM to converge to suboptimal regions 
with spuriously high curvature in the $k_b$ direction. 
As a result, empirical variance is much larger (empirical coverage $86\%$), 
while VAE provides more stable estimates and uncertainty, especially for $k_b$.

The scenario 3 lead to more complex setting: adding $k_{ac}$ 
strengthens the positive feedback $a \!\uparrow \Rightarrow c \!\uparrow,\, m \!\uparrow \Rightarrow a \!\uparrow$, 
increasing effective stiffness and deepening nonconvexity. VAE remains stable with acceptable bias and coverage, 
whereas SAEM loses practicial identifiability, 
consistent with well-known challenges of SAEM on stiff, 
highly nonlinear NLME-ODEs when curvature is poorly conditioned. 
We do not execute SAEM over the 100 simulated datasets in this scenario.

Overall, these experiments confirm that the proposed VAE approach 
can handle the nonlinear, partially stiff dynamics of the TGF-$\beta$ activation 
with reasonable accuracy and calibrated uncertainty, even when SAEM becomes numerically unstable.
\begin{table}[!htbp]
\centering
\caption{Comparison of parameter estimates on the simulation datasets from dynamics of TGF-$\beta$ activation in section \ref{sec:asthme_simu} obtained with the VAE and SAEM methods.}
\resizebox{\linewidth}{!}{%
\begin{tabular}{cccrrrrrr}
\toprule
\textbf{Scenario} & \textbf{Parameter} & \textbf{Method} &
\multicolumn{1}{c}{\shortstack{\textbf{Rel. Bias}\\($1\%$)}} &
\multicolumn{1}{c}{\shortstack{\textbf{RRMSE}\\($1\%$)}} &
\multicolumn{1}{c}{\shortstack{\textbf{Emp. Var.}\\($10^{-2}$)}} &
\multicolumn{1}{c}{\shortstack{\textbf{Est. Var.}\\($10^{-2}$)}} &
\textbf{Emp. Cov.} & \textbf{Est. Cov.} \\
\midrule
S1 & log($k_p$)      & VAE & -0.27 & 4.89 & 0.0047 & 0.0061 & 0.95 & 0.97 \\
    &            & SAEM & -2.61 & 5.47 & 0.0045 & 0.0052 & 0.90 & 0.92 \\[4pt]
    & $\omega_{k_p}$    & VAE & $\sim 0.0$ & 7.68 & 0.0015 & 0.0040 & 0.95 & 0.97 \\
    &                   & SAEM & -4.45 & 10.60 & 0.0023 & 0.0030 & 0.94 & 0.96 \\[4pt]
    &$\sigma_{a}$     & VAE & 1.65 & 3.12 & 0.07 & 0.07 & 0.94 & 0.94 \\
    &                     & SAEM & 1.75 & 3.33  & 0.08 & 0.08 & 0.95 & 0.94 \\
\midrule
S2 & log($k_p$)      & VAE & 0.37 & 6.21 & 0.010 & 0.009 & 0.96 & 0.95 \\
    &            & SAEM & 0.97 & 6.40 & 0.008 & 0.006 & 0.90 & 0.90 \\[4pt]
    & $k_b$      & VAE & 0.03 & 1.29 & 0.010 & 0.019 & 0.94 & 0.96 \\
    &            & SAEM & 1.40 & 1.90 & 0.020 & 0.0002 & 0.90 & 0.02 \\[4pt]
    & $\omega_{k_p}$    & VAE & -0.07 & 9.44 & 0.002 & 0.004 & 0.94 & 0.97 \\
    &                   & SAEM & -1.75 & 9.99 & 0.002 & 0.003 & 0.93 & 0.98 \\[4pt]
    &$\sigma_{a}$     & VAE & 1.55 & 3.07 & 0.07 & 0.07 & 0.94 & 0.94 \\
    &                     & SAEM & 1.45 & 3.18  & 0.08 & 0.08 & 0.95 & 0.94 \\
\midrule
S3 & log($k_p$)      & VAE & 1.12 & 6.48 & 0.01 & 0.009 & 0.94 & 0.93 \\[4pt]
    &$k_b$     & VAE &  1.45 & 2.56 & 0.07 & 0.1 & 0.94 & 0.96 \\[4pt]
    &log($k_{ac}$)     & VAE &  0.29 & 0.50 & 0.04 & 0.09 & 0.89 & 0.97 \\[4pt]
    & $\omega_{k_p}$  & VAE & -1.85 & 9.20 & 0.002 & 0.005 & 0.94 & 0.95 \\[4pt]
    &$\sigma_{a}$     & VAE & 1.55 & 3.07 & 0.07 & 0.07 & 0.94 & 0.94 \\
\bottomrule
\end{tabular}
}
\label{tab:asthme_results}
\end{table}
 
\section{Application on real data}
\label{sec:real_data_appli}
In the real data case, we relied on data from a cohort of Bnt162b2 vaccine recipients, in which both antibody kinetics 
and neutralizing activity were measured longitudinally against SARS-CoV-2. In brief, SARS-CoV-2 naive patients were recruited in Orléans, France
between August 27, 2020 and May 24, 2022. Individuals were followed for up to 483 days after
their first vaccine injections (see more details on the data in \citep{Planas2021}). The data pre-treatment and selection process 
is exactly the same as in \cite{Clairon2023}.
We used only the antibody kinetics data in the original dataset and analysed
$N = 25$ subjects who received three injections. 
The evolution of antibody titers (BAU/ml) is shown in Figure \ref{fig:all_data}(d)
over a 500-day period. For the analysis of this dataset, we employed exactly the same 
model design as the simulation study in the section \ref{sec:simu_antibody}.
We estimate the following parameter set:
$
\phi = \{ \log(\vartheta), \log(\bar{f}_{M_2}), \log(\bar{f}_{M_3}), \log(\delta_S),
 \log(\delta_{Ab}), \omega_\vartheta, \omega_{\bar{f}_{M_2}},\sigma_\epsilon\}, 
$
\subsection{Results}
Table~\ref{tab:real_antibody_results} reports the estimates on the real dataset obtained with both methods. 
Overall, the two methods provide consistent estimates for the population parameters and observation noise, 
with all corresponding 95\% confidence intervals overlapping. 
The main difference appears in the random-effect variance associated with $\bar{f}_{M_2}$, 
where the VAE estimate is smaller than that obtained by SAEM. 
Nevertheless, the confidence intervals still overlap, indicating no statistically significant discrepancy.
A plausible explanation is limited identifiability of the random-effect variance for $\bar{f}_{M_2}$ in the real dataset. In such
settings, different inference schemes may yield similar population means while producing more variable estimates of variance
components. The smaller VAE estimate is consistent with the well-known tendency of variational inference to underestimate posterior
uncertainty, but the overlapping confidence intervals suggest that the data do not strongly distinguish between the two estimates.
\begin{table}[!htbp]
    \centering
    \caption{VAE and SAEM parameter estimates and 95\% CIs on real dataset}
    \resizebox{\linewidth}{!}{%
    \begin{tabular}{lcccccccc}
    \toprule
    \multirow{2}{*}{\textbf{Parameter}} &
    \multicolumn{4}{c}{\textbf{VAE}} &
    \multicolumn{4}{c}{\textbf{SAEM}} \\
    \cmidrule(lr){2-5} \cmidrule(lr){6-9}
     & Est. & 95\%~CI~Lower & 95\%~CI~Upper &
       & Est. & 95\%~CI~Lower & 95\%~CI~Upper &  \\
    \midrule
    $\log(\vartheta)$      & 3.68 & 3.56 & 3.81 &  & 3.62 & 3.05 & 4.19 &  \\[4pt]
    $\log(\bar f_{M_2})$ & 1.36 & 1.32 & 1.40 &  & 1.35 & 1.01 & 1.69 &  \\[4pt]
    $\log(\bar f_{M_3})$ & 2.51 & 2.48 & 2.54 &  & 2.62 & 2.31 & 2.93 &  \\[4pt]
    $\log(\delta_S)$     & -4.72 & -4.74 & -4.70 &  & -4.72 & -4.92 & -4.51 &  \\[4pt]
    $\log(\delta_{Ab})$  & -1.89 & -2.09 & -1.69 &  & -1.91 & -2.61 & -1.21 &  \\[4pt]
    $\omega_{\vartheta}$    & 0.47 & 0.32 & 0.61 &  & 0.46 & 0.34 & 0.58 &  \\[4pt]
    $\omega_{\bar f_{M_2}}$ & 0.07 & 0.05 & 0.08 &  & 0.17 & 0.04 & 0.31 &  \\[4pt]
    $\sigma_\epsilon$           & 0.22 & 0.21 & 0.23 &  & 0.21 & 0.19 & 0.23 &  \\
    \bottomrule
    \end{tabular}
    }
\label{tab:real_antibody_results}
\end{table}    
As shown in Figure \ref{fig:Figure_antibody_identifiability}(C), the VAE and SAEM yielded consistent parameter estimates regardless of starting points.  

\begin{figure}[!htbp]
\centering

\begin{subfigure}[t]{0.49\linewidth}
    \centering
    \includegraphics[width=\linewidth]{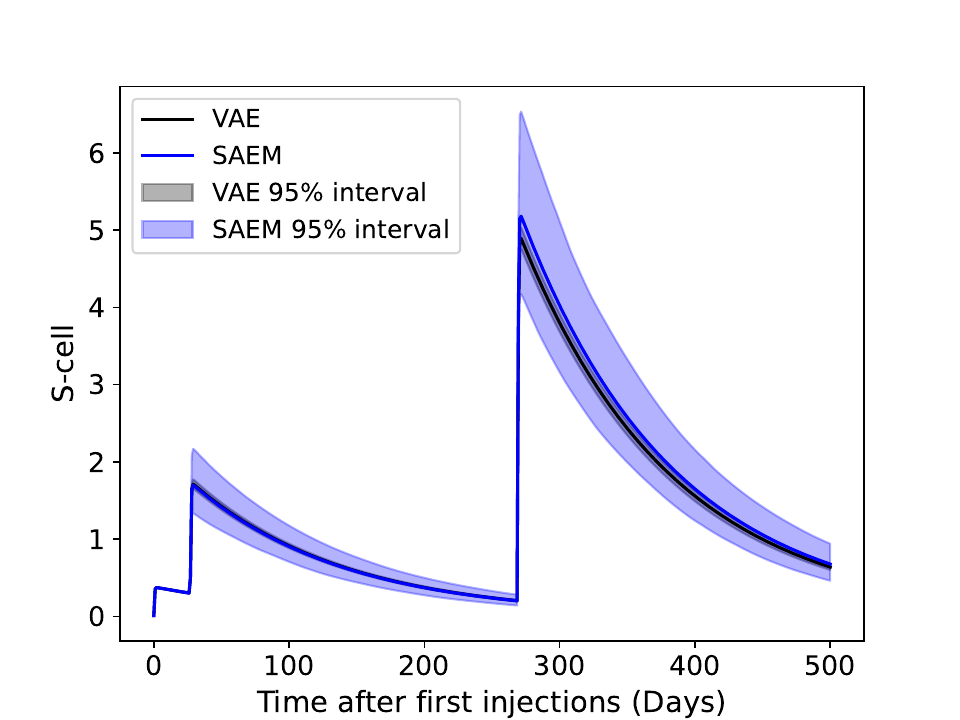}
    \caption{S-cell predictions}
    \label{fig:mean_S}
\end{subfigure}
\hfill
\begin{subfigure}[t]{0.49\linewidth}
    \centering
    \includegraphics[width=\linewidth]{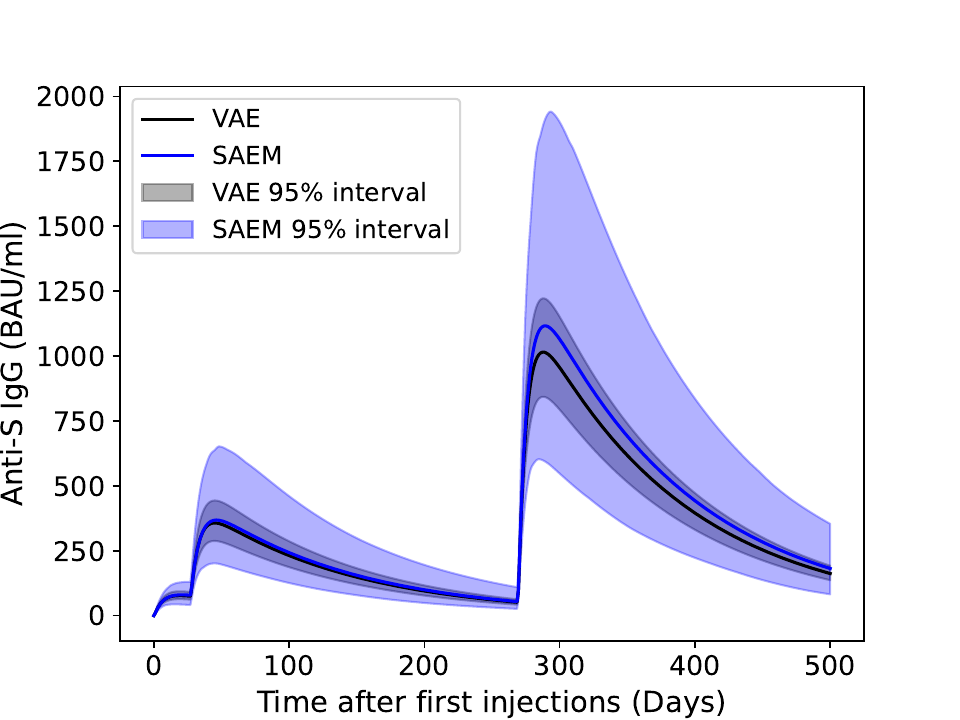}
    \caption{Anti-S IgG predictions}
    \label{fig:mean_Ab}
\end{subfigure}
\vspace{6pt}
\includegraphics[width=\linewidth]{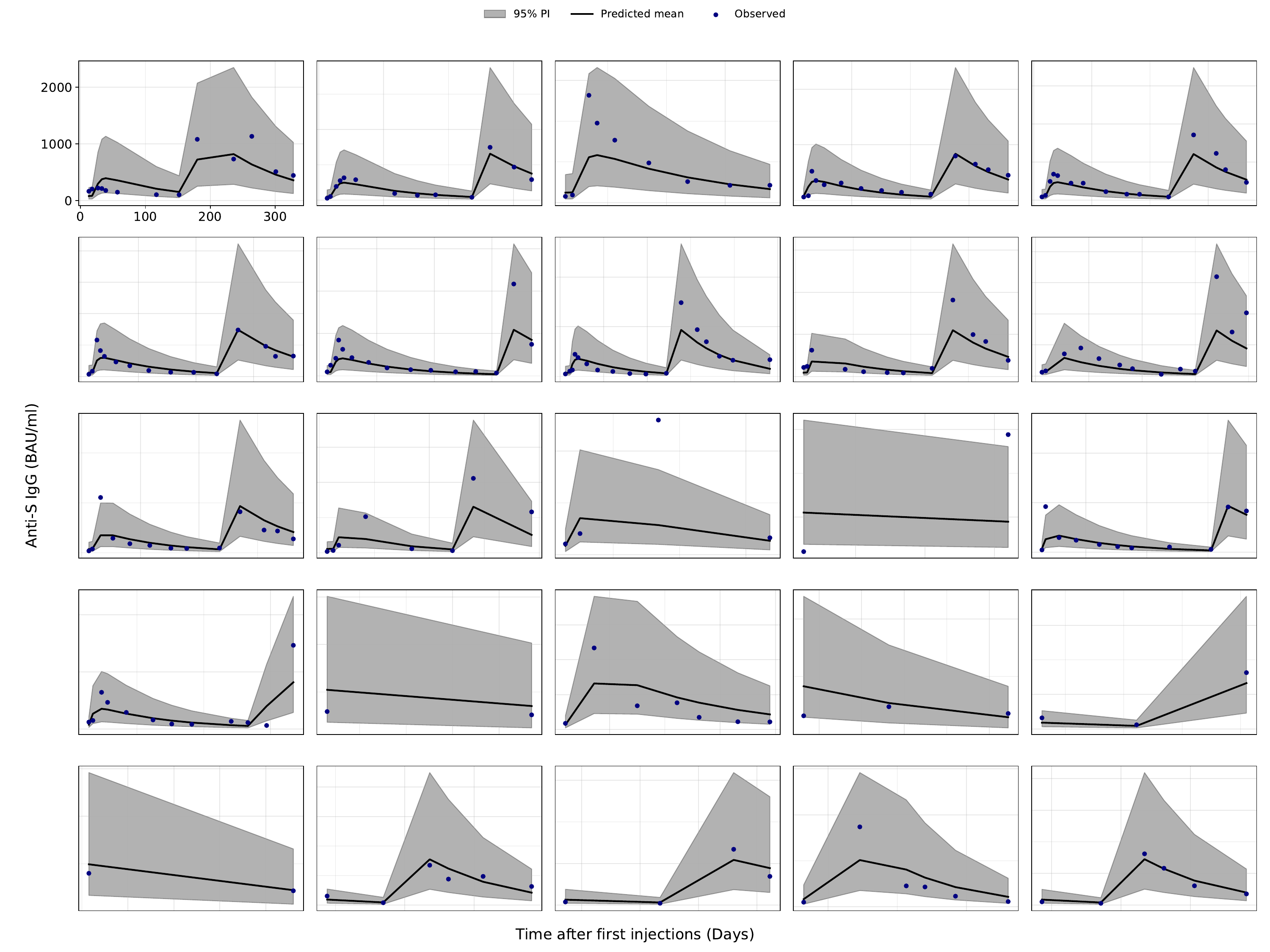}

\caption{Real-world dataset. Top: population-level predictions for (a) S-cell and (b) anti-S IgG trajectories, with 95\% uncertainty bands for VAE and SAEM.
Bottom: subject-specific predictions from the VAE with 95\% bands; points denote observed measurements.}
\label{fig:real_data_prediction}
\end{figure}

At the population level without accounting for interindividual random effects, 
Figure \ref{fig:real_data_prediction} compares the estimated population mean trajectories 
obtained by the VAE and SAEM approaches. For the antibody kinetic, 
the VAE-based trajectory provides a more accurate reconstruction: the SAEM trajectory consistently falls 
within the 95\% prediction interval of the VAE approach. 
This indicates that the variational approach captures the underlying uncertainty more faithfully, 
while SAEM remains well-aligned but more dispersed.
For the secreting cells, however, a small discrepancy appears in the peak region after 3rd injection, 
where SAEM deviates outside the VAE related confidence interval.
Figure \ref{fig:real_data_prediction} also illustrates individual predictive trajectories under the VAE-based NLME–ODEs, generated 
using the mode of the variational posterior for the subject-specific random effects $b_i$.
The predicted means match the observed data, and the 95\% prediction intervals show good overall coverage, 
confirming that the variational posterior captures the main sources of individual uncertainty. 

\section{Discussion}
In this work, we use a variational autoencoder for parameter inference in NLME–ODE models, 
replacing the sampling step in classical methods such as SAEM. 
The random-effects posterior is approximated by a Gaussian variational distribution 
parameterized by a lightweight amortized encoder, designed for sparse biological and clinical datasets. 
Throughout the design, particular attention 
was devoted to practical identifiability to ensure that the introduction of variational parameters 
does not compromise the interpretability of 
the model. For uncertainty quantification, we evaluate the observed FIM via direct hessian of 
Monte Carlo approximated marginal likelihood using
prior-based reparametrization and automatic differentiation.
In terms of estimation accuracy, for fixed-effect parameters, our variational method achieved performance 
comparable to SAEM on real data 
and outperformed it in more challenging simulated scenarios involving 
stronger nonlinearities, partial observation. 
As the number of population parameters increases, the performance gap widened in favor of the variational approach, 
illustrating its robustness to the increased curvature and multimodality of the likelihood surface—conditions 
under which SAEM often suffers from slow convergence or attraction to suboptimal local maxima.
Both the empirical coverage and the coverage predicted from the estimated variances remained generally close to the nominal 95\% level, 
especially for the fixed effect parameters.

Future extensions could focus on increasing the flexibility and robustness of the variational approximation. 
One direction is to enrich the variational posterior, through for example normalizing flows, 
to alleviate underestimation of random-effect variances and better capture posterior correlations. 
Another avenue is to move toward a fully variational Bayes formulation by placing priors on the population-level parameters.
Beyond this, we will apply the framework to higher-dimensional mechanistic models with covariates, censored data, or large sets of random effects. 
Extending the methodology to semi-parametric, for instance via neural augmentations of ODE components, 
also raises new identifiability challenges whose interaction with variational regularization deserves careful study. 
Finally, a theoretical analysis of the estimator, encompassing conditions for consistency and asymptotic normality of the ELBO estimator
would strengthen the statistical foundations of the proposed approach.

\section*{Acknowledgments}
This work receives financial support from Inserm to the booster program Exposome, also has been supported by SOLVE, 
funded by the European Union’s Horizon Europe Research and Innovation programme under grant n°101137185.
We are deeply thankful to Delphine Planas, Timothée Bruel, Laurent Hocqueloux, Thierry Prazuck and Olivier Schwartz 
for the production and the public availability of the data used in section \ref{sec:real_data_appli}. 
The experiments presented in this
paper were carried out using the PlaFRIM experimental testbed, supported by Inria, CNRS
(LABRI and IMB), Université de Bordeaux, Bordeaux INP and Conseil Régional d’Aquitaine
(see https://www.plafrim.fr). We thank Simulations Plus, Lixoft division for the free academic
use of the MonolixSuite.

\section*{SUPPLEMENTARY MATERIAL}
\begin{description}
\item Tables and Figures for Sections \ref{sec:simulation} and \ref{sec:real_data_appli} are available with this paper.
\item Code and data are available at \href{https://github.com/AuroraFr/NN_NLME/tree/paper}{https://github.com/AuroraFr/NN\_NLME/tree/paper}
\end{description}









\backmatter





%
\bibliography{NLME_biblo}
\bibliographystyle{abbrvnat} 










\label{lastpage}

\end{document}


\def\spacingset#1{\renewcommand{\baselinestretch}%
{#1}\small\normalsize} \spacingset{1}
\spacingset{1.5} 
\phantomsection\label{supplementary-material}
\bigskip

\begin{center}

{\large\bf SUPPLEMENTARY MATERIAL}

\end{center}


\section{ELBO formulation}
\begin{equation}
    \label{eq:ELBO_i}
    \mathcal{L}_{\text{elbo}_i}(\phi,\psi; Y_i)
    = \underbrace{\mathbb{E}_{q_{\psi}(b_i\mid Y_i)}\!\left[\log p_{\phi}(Y_i\mid b_i)\right]}_{\text{data fidelity}}
    \;-\;
    \underbrace{\mathrm{KL}\!\left(q_{\psi}(b_i\mid Y_i)\,\Vert\, p_{\phi}(b_i)\right)}_{\text{regularization}}.
\end{equation}
Assuming the independence between the subjects, ELBO objective function is developped as following:
\[\ln p_\phi (Y) = \sum_{i=1}^{n} \ln \int_{\mathbb{R}^{D_b}} p_\phi (Y_i | b_i) \frac{p_\phi (b_i)}{q_\psi (b_i | Y_i)}q_\psi (b_i | Y_i) db_i \]
\[= \sum_i \ln \mathbb{E}_{b_i \sim q_\psi ( \cdot | Y_i )} \left[ p_\phi (Y_i | b_i) \frac{p_\phi (b_i)}{q_\psi (b_i |Y_i)} \right] \]
\[\geq \sum_i \mathbb{E}_{b_i \sim q_\psi ( \cdot | Y_i )} \left[ \ln p_\phi (Y_i | b_i) \right] + \mathbb{E}_{b_i \sim q_\psi (\cdot | Y_i )} \left[ \ln \frac{p_\phi (b_i)}{q_\psi (b_i | Y_i)} \right] \]
\[= \sum_i \mathbb{E}_{b_i \sim q_\psi ( \cdot | Y_i )} \left[ \ln p_\phi (Y_i | b_i) \right] - KL ( q_\psi ( \cdot | Y_i) ||p_\phi ( \cdot ) ) \]
\[\approx \sum_i \frac{1}{L}\sum_{l=1}^{L} \left[ \ln p_\phi (Y_i | b^{(l)}_i) \right] - KL( q_\psi ( \cdot | Y_i) ||p_\phi ( \cdot ) ) \]
\label{appendix: ELBO_formulation}
\paragraph*{Data fidelity term}
The first term in the ELBO  given in Equation \ref{eq:ELBO_i} is the expected log-likelihood under the approximate posterior. 
It is approximed via Monte-Carlo procedure by drawing samples $b_i$ from $q_\psi$ and averaging the results. 
It can be made differentiable with respect to all its parameters ($\phi$ and $\psi$) 
using techniques like the reparameterization trick:
\begin{equation}
    b_i^{(\ell)} = \mu_\psi(Y_i) + L_{\psi}(Y_i)\varepsilon^{(\ell)}, \quad p_{\varepsilon^{(\ell)}} \sim \mathcal{N}(0, I_{D_b}).
\end{equation}
So, each sampled random effect $b_i^{(\ell)}$ induces a realization of the 
subject-specific parameter through the mapping:
$
    \theta_i^{(\ell)} =  g(\theta, b_i^{(\ell)}).
$
For each Monte-Carlo sample 
$\ell$, the ODEs defining the latent trajectory $X_i^{(\ell)}$ is solved with parameters 
$\theta_i^{(\ell)}$.
Given $X_i^{(\ell)}$, the likelihood of observed data $Y_{ij}$ is
\[
    Y_{ij} \sim p_\phi\bigl(Y_{ij}\mid b^{(l)}_i \bigr)
    = 
    \mathcal N\!\left(Y_{ij} -
        h_{\theta_i^{(\ell)}}\!\left( X_i^{(\ell)}(t_{ij}) \right),
        \Sigma
    \right),
    \quad j = 1,\dots,n_i.
\]
Thus, the expected log-likelihood term appearing in the ELBO is approximated by
\begin{align}
    \mathbb{E}_{q_\psi(b_i \mid Y_i)}\!\left[\log p_\phi(Y_i \mid b_i)\right]
    \approx \frac{1}{N_{MC}}\sum_{\ell=1}^{N_{MC}}\sum_{j=1}^{n_i}
       \log p_\phi\bigl(Y_{ij}\mid h_{\theta_i^{(\ell)}}(X_i^{(\ell)}(t_{ij})),\Sigma\bigr)
    \notag\\
    \propto 
    -\frac{n_i}{2}
    \sum_r \log \sigma_{\varepsilon,r}^2
    -\frac{1}{2N_{MC}} 
    \sum_{\ell=1}^{N_{MC}} 
    \sum_{j=1}^{n_i}
    \big(Y_{ij} - h_{\theta_i^{(\ell)}}(X_i^{(\ell)}(t_{ij}))\big)^{\!\top}
    \Sigma^{-1}
    \big(Y_{ij} - h_{\theta_i^{(\ell)}}(X_i^{(\ell)}(t_{ij}))\big).
\end{align}
Here, $N_{MC}$ denotes the number of samples drawn from the standard gaussian distribution.
A larger $N_{MC}$ generally improves the accuracy of the ELBO estimation and reduces the variance of the stochastic gradient, 
but it also increases the computational cost per iteration proportionally because $X^{(l)}_i(t)$ depends on $b^{(l)}_i$ through the ODEs, this expectation requires solving the ODEs for each sample $b^{(l)}_i$.
In practice, moderate values of $N_{MC}$(e.g., 10–50) often provide a good trade-off between estimation precision and runtime efficiency. 

\paragraph*{KL divergence term}
The second term is the Kullback--Leibler divergence between the variational distribution $q_{\psi}$
and the prior distribution of the random effects $p_{\phi}$, which penalizes deviations from the
population-level distribution and thus controls the complexity of the learned individual posteriors.
This term admits a closed-form expression in our Gaussian setting. We consider the full-covariance case
and parameterize both covariance matrices via their Cholesky factors:
\[
p_\phi(b_i)=\mathcal N(\mathbf 0,\Omega),\qquad \Omega=L_\Omega L_\Omega^\top,
\]
\[
q_\psi(b_i\mid Y_i)=\mathcal N(\mu_{\psi}(Y_i),\Sigma_{\psi}(Y_i)),\qquad \Sigma_{\psi}(Y_i)=L_{\psi}(Y_i)L_{\psi}(Y_i)^\top,
\]
where $L_\Omega$ and $L_{\psi}(Y_i)$ are lower-triangular with positive diagonal entries and $D_b=\dim(b_i)$.

The Expected log-prior is:
\[
\begin{aligned}
\mathbb E_{q_\psi(b_i\mid Y_i)}[\log p_\phi(b_i)]
&=
\int q_\psi(b_i\mid Y_i)\,\log \mathcal N(\mathbf 0,\Omega)\,db_i \\
&=
-\frac{D_b}{2}\log(2\pi)
-\frac{1}{2}\log\det\Omega
-\frac{1}{2}\Big(
\mu_{\psi}(Y_i)^\top \Omega^{-1}\mu_{\psi}(Y_i)
+\mathrm{tr}(\Omega^{-1}\Sigma_{\psi}(Y_i))
\Big).
\end{aligned}
\]

The Expected log-variational density (negative entropy) is: 
\[
\begin{aligned}
\mathbb E_{q_\psi(b_i\mid Y_i)}[\log q_\psi(b_i\mid Y_i)]
&=
\int q_\psi(b_i\mid Y_i)\,\log q_\psi(b_i\mid Y_i)\,db_i \\
&=
-\frac{D_b}{2}\log(2\pi)
-\frac{1}{2}\log\det\Sigma_{\psi}(Y_i)
-\frac{D_b}{2}.
\end{aligned}
\]

Therefore,
\[
\begin{aligned}
\mathrm{KL}\!\left(q_\psi(b_i\mid Y_i)\,\|\,p_\phi(b_i)\right)
&=
\frac{1}{2}\Big(
\mathrm{tr}(\Omega^{-1}\Sigma_{\psi}(Y_i))
+\mu_{\psi}(Y_i)^\top \Omega^{-1}\mu_{\psi}(Y_i)
-D_b
+\log\frac{\det\Omega}{\det\Sigma_{\psi}(Y_i)}
\Big).
\end{aligned}
\]

and the Log-determinants via Cholesky factors:
\[
\log\det\Omega = 2\sum_{k=1}^{D_b}\log (L_{\Omega,kk}),
\qquad
\log\det\Sigma_{\psi}(Y_i) = 2\sum_{k=1}^{D_b}\log (L_{\psi,kk}(Y_i)).
\]

\section{Uncertainty quantification}
To evaluate $p_{\hat{\phi}}(Y_i)$ and its derivatives, we apply a Monte Carlo approximation
based on the reparameterization trick.
We express the random effects $b_i$ as a transformation
of a noise variable $p_{\epsilon} \sim \mathcal{N}(\mathbf{0}, I_{D_b})$ independent of $\phi$:
$b_i = \mathcal{T}_{\hat{\phi}}(\epsilon) = \hat{\Omega}^{1/2}\epsilon$
where $\hat{\Omega}^{1/2}$ is the estimated standard deviation matrix
of the prior distribution $p_{\hat{\phi}}(b_i)$ which is assumed to be a zero-mean Gaussian. So we can
approximate the marginal likelihood using monte carlo as:
\begin{equation}
p_{\hat{\phi}}(Y_i) \approx 
\frac{1}{L}\sum_{\ell=1}^{L} p_{\hat{\phi}}\bigl(Y_{i}\mid b_i^{(l)} \bigr) 
= \frac{1}{L}\sum_{\ell=1}^{L} p_{\hat{\phi}}\bigl(Y_{i}\mid \mathcal{T}_{\hat{\phi}}(\epsilon^{(l)}) \bigr),
\qquad p_{\epsilon^{(l)}} \sim \mathcal{N}(\mathbf{0}, I_{D_b}).
\end{equation}
The corresponding gradient and Hessian of the marginal likelihood are:
\[
\nabla_{\phi} p_{\hat{\phi}}(Y_i)
 = \mathbb{E}_{p(\epsilon)}\big[\nabla_{\phi} p_{\hat{\phi}}(Y_i \mid \mathcal{T}_{\hat{\phi}}(\epsilon))\big]
  \approx \frac{1}{L}\sum_{\ell=1}^{L} \nabla_{\phi} p_{\hat{\phi}}(Y_i \mid \mathcal{T}_{\hat{\phi}}(\epsilon^{(l)})),\\[3pt]
 \]
 \[ 
\nabla_{\phi}^2 p_{\hat{\phi}}(Y_i)
 = \mathbb{E}_{p(\epsilon)}\big[\nabla_{\phi}^2 p_{\hat{\phi}}(Y_i \mid \mathcal{T}_{\hat{\phi}}(\epsilon))\big]
  \approx \frac{1}{L}\sum_{\ell=1}^{L} \nabla_{\hat{\phi}}^2 p_{\hat{\phi}}(Y_i \mid \mathcal{T}_{\hat{\phi}}(\epsilon^{(l)})).
\]
Now,  we can rely on previously used gradient backpropagation techniques to efficiently appproximates $\nabla_{\phi} p_{\hat{\phi}}(Y_i)$
and $\nabla_{\phi}^2 p_{\hat{\phi}}(Y_i)$ thanks to the identities:
\[
p_{\hat{\phi}}(Y_i \mid \mathcal{T}_{\hat{\phi}}(\epsilon)) = \exp\left( \log p_{\hat{\phi}}(Y_i \mid \mathcal{T}_{\hat{\phi}}(\epsilon)) \right),
\]
\[
\nabla_\phi p_{\hat{\phi}}(Y_i \mid \mathcal{T}_{\hat{\phi}}(\epsilon)) = \nabla_\phi \log p_{\hat{\phi}}(Y_i \mid \mathcal{T}_{\hat{\phi}}(\epsilon)) \cdot p_{\hat{\phi}}(Y_i \mid \mathcal{T}_{\hat{\phi}}(\epsilon)),
\]
\[
\nabla^2_\phi p_{\hat{\phi}}(Y_i \mid \mathcal{T}_{\hat{\phi}}(\epsilon)) = \left( \nabla^2_\phi \log p_{\hat{\phi}}(Y_i \mid \mathcal{T}_{\hat{\phi}}(\epsilon)) + \nabla_\phi \log p_{\hat{\phi}}(Y_i \mid \mathcal{T}_{\hat{\phi}}(\epsilon)) \nabla_\phi \log p_{\hat{\phi}}(Y_i \mid \mathcal{T}_{\hat{\phi}}(\epsilon))^\top \right) \cdot p_{\hat{\phi}}(Y_i \mid \mathcal{T}_{\hat{\phi}}(\epsilon)).
\]
Finally, the Hessian of the log-marginal likelihood is computed as:
\begin{equation}
\nabla^2_\phi \log p_{\hat{\phi}}(Y_i) = \frac{\nabla^2_\phi p_{\hat{\phi}}(Y_i) \cdot p_{\hat{\phi}}(Y_i) - \nabla_\phi p_{\hat{\phi}}(Y_i) \nabla_\phi p_{\hat{\phi}}(Y_i)^\top}{p_{\hat{\phi}}(Y_i)^2}.
\end{equation}
\section{Solving the ODE of antibody kinetic and Identifiability Implications}
\label{appendix:antibody_structural_identifiability}

We consider the following linear system of ODEs modeling antibody dynamics:
\[
\begin{cases}
\dot{S}(t) = \bar{f}_M(t) e^{-\delta_V (t - t_k)} - \delta_S (t)\\
\dot{Ab}(t) = \vartheta S(t) - \delta_{Ab} Ab(t) \\
(S(0), Ab(0)) = (0.01, 0.1) 
\end{cases}
\]

Step 1: Solving the equation for $S(t)$
The ODE for \(S(t)\) is linear and non-homogeneous. We use the integrating factor method.
\paragraph*{Integrating factor:}
\[
\mu(t) = e^{\delta_S t}
\]
Multiply both sides:
\[
e^{\delta_S t} \dot{S}(t) + \delta_S e^{\delta_S t} S(t) = e^{\delta_S t} \bar{f}_M(t) e^{-\delta_V (t - t_k)}
\Rightarrow
\frac{d}{dt} \left( e^{\delta_S t} S(t) \right) = e^{\delta_S t} \bar{f}_M(t) e^{-\delta_V (t - t_k)}
\]
Integrate both sides:
\[
e^{\delta_S t} S(t) = \int_0^t \bar{f}_M(\tau) e^{-\delta_V(\tau - t_k)} e^{\delta_S \tau} \, d\tau
\Rightarrow
S(t) = \int_0^t \bar{f}_M(\tau) e^{-\delta_V(\tau - t_k)} e^{-\delta_S (t - \tau)} \, d\tau
\]

\paragraph*{Final result:}
\[
\boxed{
S(t) = \left(\bar{f}_M(\cdot) e^{-\delta_V(\cdot - t_k)} \right) * e^{-\delta_S t}
}
\]
This is a convolution of the input stimulus with an exponential decay kernel governed by \(\delta_S\).

\paragraph*{Step 2: Solving the equation for $Ab(t)$}
The equation for \(Ab(t)\) is also linear:
\[
\dot{Ab}(t) = \vartheta S(t) - \delta_{Ab} Ab(t)
\Rightarrow
Ab(t) = \int_0^t \vartheta S(\tau) e^{-\delta_{Ab}(t - \tau)} \, d\tau
= \vartheta \cdot (S * e^{-\delta_{Ab} t})
\]
Substitute the expression for \(S(t)\):

\[
Ab(t) = \vartheta \cdot \left( \left(\bar{f}_M(\cdot) e^{-\delta_V(\cdot - t_k)} \right) * e^{-\delta_S t} * e^{-\delta_{Ab} t} \right)
\]
By associativity of convolution:
\[
\boxed{
Ab(t) = \vartheta \cdot \left(\bar{f}_M(t) e^{-\delta_V(t - t_k)} * h(t)
\right), \quad \text{where } h(t) = e^{-\delta_S t} * e^{-\delta_{Ab} t}
}
\]

\paragraph*{Interpretation: Convolution-Like Structure and Parameter Symmetry}

The function \(Ab(t)\) is obtained by a chain of linear filters (convolutions).
First, the exogenous input \(f_M(t) e^{-\delta_V(t - t_k)}\) is smoothed by the decay of \(S(t)\) 
via convolution with \(e^{-\delta_S t}\).
Then, the antibody dynamics further smooth the result via convolution with \(e^{-\delta_{Ab} t}\).

Because convolution of exponentials is commutative:
\[
e^{-\delta_S t} * e^{-\delta_{Ab} t} = e^{-\delta_{Ab} t} * e^{-\delta_S t}
\]
The final result \(Ab(t)\) depends on the combined smoothing effect of \(\delta_S\) and \(\delta_{Ab}\), not their individual identities.
This leads to a key structural symmetry:
\[
(\delta_S, \delta_{Ab}) \quad \text{and} \quad (\delta_{Ab}, \delta_S)
\]
can yield the same \(Ab(t)\) trajectory.
Thus, the parameters \(\delta_S\) and \(\delta_{Ab}\) are Not structurally identifiable from \(Ab(t)\) alone.

\section{Optimization and Regularization}

To improve convergence and generalization in sparse data regimes, 
we include the option dropout in the encoder network to prevent overfitting.
To train the model, we maximize the ELBO using stochastic gradient-based optimization. The Adam optimizer is used, 
and all gradients are computed using automatic differentiation 
with the software \texttt{Optax}. Optionally, we can use a cosine annealing learning rate schedule 
to facilitate smoother optimization.

In the simulation studies, model fitting is performed on a training dataset, while an 
independent validation dataset is generated from the same data-generating mechanism and used to 
monitor performance and tune hyperparameters. In the real-data application, where only a single 
dataset is available, we instead construct a validation set by splitting the observed data.

To ensure numerical stability and avoid premature stopping, we monitor three complementary convergence criteria during training: 
the gradient norm, the parameter update magnitude, and the validation loss evolution.
Let $s$ denote the epoch of the optimization algorithm.

\paragraph*{Gradient norm criterion.}
Convergence toward a stationary point of the ELBO objective is assessed by the Euclidean norm 
of its gradient with respect to all model parameters $(\phi, \psi)$:
\[
\bigl\|\nabla_{(\phi,\psi)} \mathcal{L}^{(s)}_{\text{ELBO}}(\phi,\psi)\bigr\|_2
= \sqrt{
\left(
    \sum_{j=1}^{D_\phi + D_\psi}
    \left(
    \frac{\partial \mathcal{L}^{(s)}_{\text{ELBO}}(\phi,\psi)}{\partial (\phi, \psi)_j}
    \right)^{\!2}
\right)} < \varepsilon_g,
\label{eq:grad_criterion}
\]
where $\|\cdot\|_2$ denotes the Euclidean norm and 
$\varepsilon_g$ is a small positive tolerance, typically in the range $10^{-5}$--$10^{-3}$. 
This condition indicates that further optimization steps would lead to negligible 
improvement in the objective.

\paragraph*{Parameter update criterion.}
Convergence in parameter space is additionally checked by comparing successive parameter estimates:
\[
\frac{
    \bigl\|(\phi, \psi)^{(s)} - (\phi, \psi)^{(s-1)}\bigr\|_2
}{
    \bigl\|(\phi, \psi)^{(s-1)}\bigr\|_2 + \varepsilon
}
< \varepsilon_p,
\label{eq:param_update_criterion}
\]

\[
\bigl\|(\phi, \psi)^{(s)} - (\phi, \psi)^{(s-1)}\bigr\|_2
=
\sqrt{\left(
    \sum_{j=1}^{D_\phi + D_\psi}
    \bigl((\phi, \psi)^{(s)}_j - (\phi, \psi)^{(s-1)}_j\bigr)^{2}
\right)},
\]
and $\varepsilon_{p}$ is a small threshold (typically $10^{-7}$–$10^{-5}$) and $\varepsilon$ (typically $10^{-8}$) prevents division by zero.
This criterion ensures that the parameter updates have become sufficiently small.

\paragraph*{Validation loss criterion.}
Finally, an early stopping rule is applied based on the validation ELBO loss 
$\mathcal{L}^{(s)}_{\mathrm{val}}$:
\[
\label{eq:val_loss_criterion}
\mathcal{L}^{(s)}_{\mathrm{val}}
>
\min_{s^{\prime} < s} \mathcal{L}^{(s^{\prime})}_{\mathrm{val}}
\quad
\text{for more than } P \text{ consecutive epochs,}
\]
where $P$ is a predefined patience parameter.
This prevents overfitting by stopping training once the validation performance
no longer improves.

\section{Simulation details}
To ensure positivity of the variance parameters, 
the encoder outputs log-standard-deviation, i.e. it predicts $\log(\Sigma_{\psi}(Y_i))$, and 
we estimate $\log(\Sigma^{2})$ for the observation noise.

All experiments were run on a single machine equipped with an NVIDIA A100 GPU (40 GB), using float64 precision. 
Importantly, the proposed VAE-based NLME-ODE approach is not GPU-intensive: 
we were able to run the full pipeline on a standard laptop (Intel Core i7 CPU with an NVIDIA RTX 2000 GPU) including the variance estimation step, 
though it is slower than on the A100. The pipeline can also be run on CPU-only hardware with the same workflow and is slower still.

For the SAEM benchmark, we used \texttt{Monolix 2023R1} on a high-performance CPU server equipped 
with 2x 16-core Skylake Intel 
Xeon Gold 6142 @ 2.6 GHz and 384 GB of RAM. 
For most models, Monolix was run with default settings, which include using \texttt{autoChains = TRUE}, 
meaning Monolix automatically determines the number of MCMC chains (starting from one) based on internal convergence diagnostics.
The simulated annealing option is enabled which permits to keep the explored parameter space large for a longer time.
For the complex settings in the antibody kinetic model and the TGF-$\beta$ model, 
we used a more robust configuration to stabilize SAEM convergence. In particular, we set the number of chains as 10 
and customized the MCMC strategy to \texttt{c(5, 2, 5)}, 
giving more iterations to the blockwise kernel, which improves mixing in these models. 
We also increased the burn-in period to 100 iterations and used default 500 exploratory iterations with auto-stop disabled. 
Finally, we set the smoothing interval to 300 and disabled automatic smoothing, which avoids premature variance shrinkage. 
\subsection{Pharmacokinetic model}
\label{simu:pkpd}
To assess the accuracy and robustness of our variational inference framework, 
we start the comparison with a classic and partially observed pharmacokinetic model with 
first-order absorption and elimination rates defined by the following subject-specific ODEs:
\begin{equation}
\begin{cases}
\label{eq:pkpd_model}
\dot{X}_{1,i}(t) = \vartheta_{2} X_{2,i}(t) - \vartheta_{1,i} X_{1,i}(t), \\
\dot{X}_{2,i}(t) = - \vartheta_{2} X_{2,i}(t), \\
(X_{1,i}(0), X_{2,i}(0)) = (2, 3),
\end{cases}
\end{equation}
where \( \vartheta_{1,i}, \vartheta_{2} \) are the parameters governing the decay and transfer rates between two compartments.
Structural identifiability of the system in Eq. \eqref{eq:pkpd_model} 
holds provided that the initial conditions are known and at least one state variable (typically $X_{1,i})$
is observed continuously over time.

\noindent The parameter \( \vartheta_{1,i} \) varies across individuals and is modeled as:
\[
\log(\vartheta_{1,i}) = \log(\vartheta) + b_i, \quad b_i \sim \mathcal{N}(0, \omega^2_{\vartheta_1}),
\]
where \( \vartheta_1 \) is the population-level mean and \( \omega_{\vartheta_1} \) is the standard deviation of the random effect.
The dataset consists of 100 subjects. For each subject $i$, we observe $6$ noisy partial observations 
of the first state $X_{1,i}$ at time points $t_{ij}$ over the interval $T=[0,10]$:
\[
Y_{ij} = X_{1,i}(t_{ij}) + \epsilon_{ij}, \quad \epsilon_{ij} \sim \mathcal{N}(0, \sigma^2_{\epsilon}).
\] The true parameter values are in the table \ref{tab:true_values_pkpd}.

\begin{table}[ht]
\centering
\begin{tabular}{lcc}
\hline
\textbf{Parameter} & \textbf{Description} & \textbf{True value} \\
\hline
$\vartheta_1$ & Elimination rate of $X_{1,i}$ (individual-specific) & $0.5$ \\
$\vartheta_2$ & Transfer rate from $X_{2,i}$ to $X_{1,i}$ & $2$ \\
$\omega_{\vartheta_1}$ & Std.~dev. of random effect on $\vartheta_1$ & $0.5$ \\
$\sigma_\varepsilon$ & Residual error standard deviation & $0.2$ \\
\hline
\end{tabular}
\caption{True parameter values used in the simulation datasets of the pharmacokinetic model}
\label{tab:true_values_pkpd}
\end{table}
In the irregular sampling setting, we used 
a single-layer Gated Recurrent Unit (GRU) with hidden state of dimension $16$ in the encoder.
The final GRU hidden state is mapped through a linear layer to a 4-dimensional embedding,
followed by a GELU nonlinearity.

In addition, we used a custom initialization for the final linear layer: 
weights were re-initialized from a normal distribution with a small standard deviation (\texttt{0.001}) 
to prevent excessively large activations and variance 
at the beginning of training.
We used an initial learning rate of 0.1. 
If the monitored metric did not improve for 50 consecutive epochs, the learning rate was reduced to 0.05. 
Training was stopped if no improvement was observed for 100 epochs. We integrated the ODEs using 
the adaptive fifth-order explicit Runge–Kutta method
(Dormand–Prince 5(4);\texttt{dopri5}) with \(\texttt{rtol}=\texttt{atol}=10^{-8}\). The average training time was approximately 20 minutes.

\begin{figure}[!htbp]
    \centering
    \includegraphics[width=0.75\textwidth]{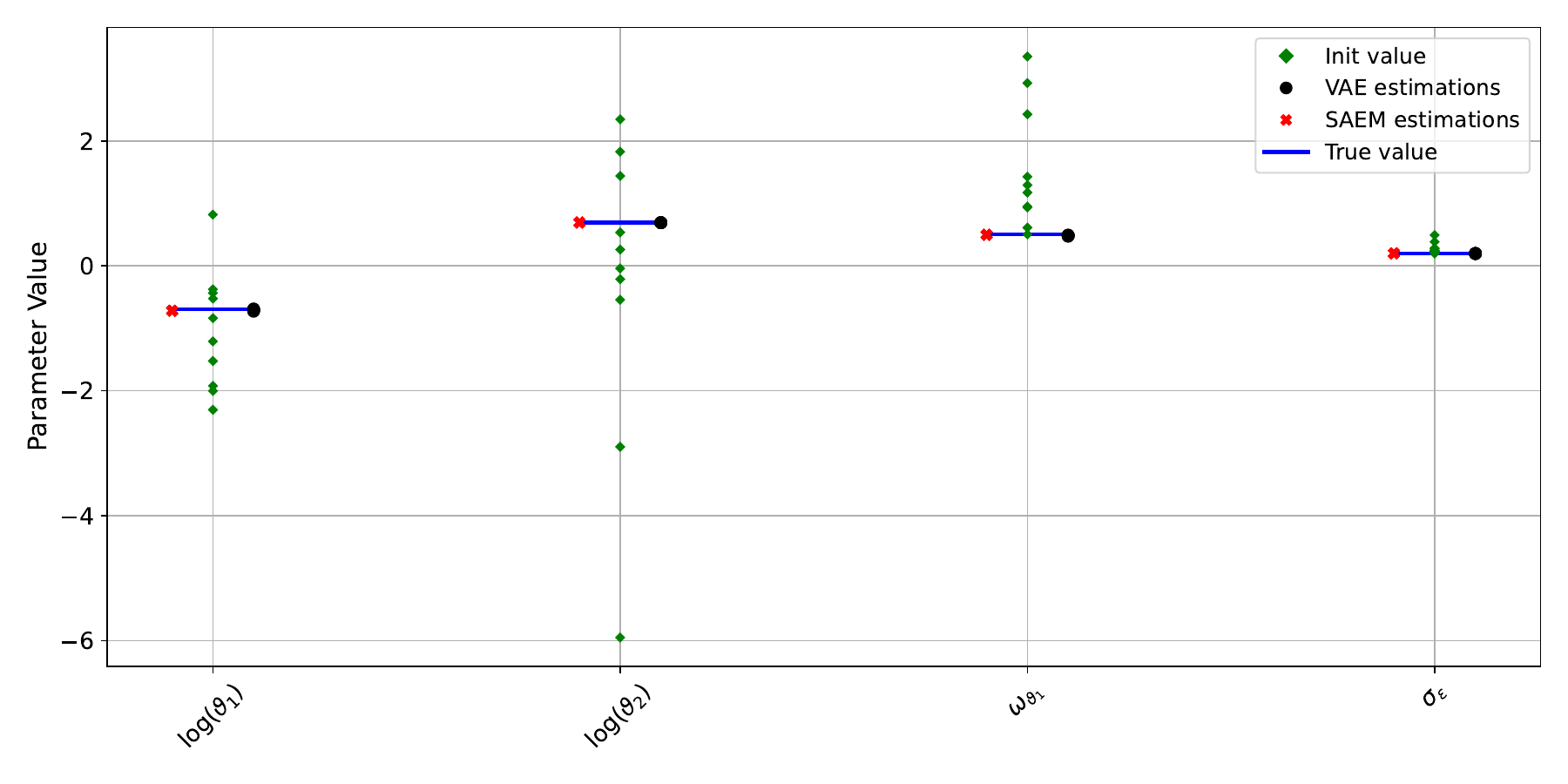}
    \caption{Convergence assessement of SAEM and VAE in pharmacokinetic model. For each parameter, we show the scatter of estimates over random initializations.
A single tight cluster around the true value indicates identifiability, whereas multiple separated clusters reveal competing local optima and lack of robustness.}
    \label{fig:Figure_pkpd_identifiability}
\end{figure}

We quantified the pointwise estimation accuracy and variance of our estimator, as summarized in Table~\ref{tab:pkpd_results}. 
Overall, both methods recover the fixed effects $\log(\vartheta_1)$ and $\log(\vartheta_2)$ well, 
as indicated by small relative biases and low RRMSE values. 
In contrast, the inter-individual variability parameter $\omega_{\vartheta_1}$, 
exhibits larger RRMSEs and relative biases than the fixed-effect parameters, 
reflecting the greater difficulty of estimating random-effect variance components from sparse per-subject sampling.
Finally, the close agreement between empirical and estimated variances and the near-nominal coverage probabilities ($\approx 0.95$) 
suggest that the uncertainty quantification provided by both methods is reasonably well calibrated in this setting.
Moreover, we verified that our method does not introduce additional practical identifiability issues in Figure \ref{fig:Figure_pkpd_identifiability}.
This first example is a simple one for which SAEM based approach is known to succeed. 
It illustrates our method's ability to retreive the same accuracy level as classic methods 
for problem of reasonable complexity.
\begin{table}[!htbp]
\centering
\caption{Comparison of parameter estimates for the pharmacokinetic model in \ref{simu:PKPD} obtained with the VAE and SAEM methods.}
\scriptsize
\begin{tabular}{ccrrrrrr}
\toprule
\textbf{Parameter} & \textbf{Method} &
\multicolumn{1}{c}{\shortstack{\textbf{Rel. Bias}\\\scriptsize($\%$)}} &
\multicolumn{1}{c}{\shortstack{\textbf{RRMSE}\\\scriptsize($\%$)}} &
\multicolumn{1}{c}{\shortstack{\textbf{Emp. Var.}\\\scriptsize($10^{-2}$)}} &
\multicolumn{1}{c}{\shortstack{\textbf{Est. Var.}\\\scriptsize($10^{-2}$)}} &
\textbf{Emp. Cov.} & \textbf{Est. Cov.} \\
\midrule
    $\log(\vartheta_1)$      & VAE & 1.04 & 6.58 & 0.20 & 0.25 & 0.95 & 0.96 \\
                         & SAEM & -0.50 & 7.24 & 0.25 & 0.25 & 0.94 & 0.94 \\[2pt]
    $\log(\vartheta_2)$     & VAE &  -0.10 & 2.63 & 0.03 & 0.04 & 0.94 & 0.95 \\
                         & SAEM & 0.04 & 2.64 & 0.03 & 0.03 & 0.95 & 0.95 \\[2pt]
    $\omega_{\vartheta_1}$     & VAE & 2.56 & 10.42 & 0.49 & 0.56 & 0.94 & 0.97 \\
                         & SAEM & 2.39 & 10.35  & 0.49 & 0.50 & 0.94 & 0.95 \\[2pt]
    $\sigma_{\epsilon}$     & VAE & -0.31 & 1.42 & 0.21 & 0.27 & 0.94 & 0.95 \\
                         & SAEM & -0.07 & 1.97  & 0.40 & 0.41 & 0.94 & 0.94 \\[2pt]
\bottomrule
\end{tabular}

\label{tab:pkpd_results}
\end{table}

\subsection{Antibody kinetics}
For the implementation of the encoder, in the regular sampling design we applied \texttt{Conv1d} 
with \texttt{kernel\_size}=3 and a latent representation of dimension $4$. 
For the irregular sampling design, we replaced the convolutional feature extractor 
with a single-layer GRU with hidden
dimension $32$ to encode the input sequence. We used an initial learning rate of 0.05. If the
monitored metric did not improve for 200 consecutive epochs, the learning rate was reduced to 0.01. 
The same solver is used as in \ref{simu:pkpd}.
Training was stopped if no improvement was observed for 500 epochs.  
The average training time was approximately 45 minutes.
\begin{table}[ht]
\centering
\begin{tabular}{lcc}
\hline
\textbf{Parameter} & \textbf{Description} & \textbf{True value} \\
\hline
$\vartheta$   & Antibody production rate & $24.5$ \\
$\bar{f}_{M_2}$   & Memory factor after 2nd injection  & $7.1$  \\
$\bar{f}_{M_3}$      & Memory factor after 3rd injection  & $18.5$ \\
$\delta_S$ & Secreting cell decay rate       & $0.01$ \\
$\delta_{Ab}$ & Antibody degradation rate                 & $0.08$ \\
$\omega_\vartheta$  & Std.~dev. of random effect on $\theta$       & $0.5$ \\
$\omega_{\bar{f}_{M_2}}$   & Std.~dev. of random effect on $\bar{f}_{M_2}$          & $0.9$ \\
$t_0$   & First injection time (Days)         & $0$ \\
$t_1$   & Second injection time (Days)           & $30$ \\
$t_2$   & Third injection time (Days)         & $250$ \\
$\epsilon_{Ab}$   & Residual error standard deviation         & $0.1$ \\
\hline
\end{tabular}
\caption{True parameter values used in the simulation datasets of antibody kinetic model}
\label{tab:true_values_antibody}
\end{table}

\subsection{TGF-$\beta$ dynamics}
The parameter values were chosen as shown in table \ref{tab:tgf_params}.
\begin{table}[ht]
\centering
\begin{tabular}{lcc}
\hline
\textbf{Parameter} & \textbf{Meaning} & \textbf{Value} \\ 
\hline
$k_p$           & Proliferation rate of ASM & 1.15 \\ 
$\omega_{k_p}$  & Std.\,dev.\,of random effect on $k_p$ & 0.05 \\ 
$k_{ac}$        & Activation rate of TGF-$\beta$ via ASM contraction & 0.01 \\ 
$k_b$           & TGF-$\beta$ binding rate to ASM receptors & 1.00 \\ 
$\phi_c$        & Contractile ASM apoptosis rate & 0.10 \\ 
$k_s$           & Amplitude of external stimulus & 0.20 \\ 
$\nu$           & Stimulus duration & 30.0 \\ 
$\tau_i$           & Stimulus times & $[50,\, 90,\, 130,\, 170,\, 210]$ \\  
$\epsilon_{a}$   & Residual error standard deviation         & $0.1$ \\
\hline
\end{tabular}
\caption{Parameter values used in the TGF-$\beta$ model simulations (diseased-state regime). 
Parameters not listed here were fixed to their baseline values reported in Table~A4 of}
\label{tab:tgf_params}
\end{table}
In this simulation, we used the same encoder architecture as in the antibody kinetic model for the regular sampling, 
but with a larger convolutional receptive field by setting \texttt{kernel\_size}=7. The L-stable
fifth-order ESDIRK method \texttt{Kvaerno5} with \(\texttt{rtol}=\texttt{atol}=10^{-6}\) is used to integrate the ODEs.
Optimization was
performed with a default learning rate of $5\times 10^{-4}$, and we applied global gradient clipping
using \texttt{optax.clip\_by\_global\_norm(1.0)} to improve numerical stability during training. 
The underlying dynamic system is moderately stiff with strongly coupled states, which can lead to ill-conditioned optimization and
occasional large gradient norms (e.g., due to sensitivity amplification through the ODE solver).
Clipping the global gradient norm prevents severe exploding updates and helps maintain steady progress 
in ELBO optimization. 
Due to the computational cost of integrating the stiff ODE system, the average training time was approximately 70 minutes. 

\subsection{Real-world antibody kinetic dataset}
In real-world applications, longitudinal sequences typically have variable lengths.
To enable efficient batching, we pad each sequence to a common length $T$
(the maximum sequence length in the dataset) and define a binary mask
$m_{it}\in\{0,1\}$ indicating whether time step $t$ for subject $i$ is observed
($m_{it}=1$) or corresponds to padding/missingness ($m_{it}=0$). 
The mask is applied throughout the encoder and in the reconstruction term, ensuring
that padded positions neither affect the pooled representation nor contribute to the objective.
We then use masked attention pooling to aggregate
variable-length sequences.
Let $z_i = (z_1,\dots,z_{T})$ be a sequence of hidden states per subject with the vector $z_t \in \mathbb{R}^{\text{H}}$
after the convolution layer where $\text{H}$ is the dimension of the hidden state.
We compute a scalar attention logit for each time step using a learned linear map:
\[
e_t = w^\top z_t + c, \qquad w \in \mathbb{R}^{\text{H}},\; c\in \mathbb{R}.
\]
Masking is enforced by assigning a large negative value to invalid positions:
\[
\tilde{e}_t =
\begin{cases}
e_t, & \text{if } m_{it}>0,\\
-10^{9}, & \text{otherwise}.
\end{cases}
\]
The attention weights are then obtained by a softmax over time step,
\[
\alpha_t = \frac{\exp(\tilde{e}_t)}{\sum_{t=1}^{T}\exp(\tilde{e}_t)}, \qquad t=1,\dots,T,
\]
so that $\sum_{t=1}^{T}\alpha_t = 1$ and masked positions receive numerically zero weight.
Finally, the pooled representation is a combination of the hidden states:
\[
\mathrm{pool}(z_i,m_i) \;=\; \sum_{t=1}^{T} \alpha_t\, e_t \in \mathbb{R}^{\text{H}}.
\]
This pooling layer produces a H-dimensional summary vector 
while allowing the model to focus on the most informative time steps 
and ignore padded/missing entries through the mask.

We padded each subject trajectory to a common length $17$ with a binary mask and applied a \texttt{Conv1d} 
with \texttt{kernel\_size}=1 and \texttt{output\_channels = 16}. The resulting padded sequence is
summarized via masked attention pooling and then
the pooled vector of $\mathbb{R}^{16}$ was transformed as an embedding of $4$-dimension.

In the training, with only $25$ subjects available, splitting the data into separate training and 
validation sets would leave too few subjects in the validation subset, leading to extremely noisy 
validation metrics and unstable hyperparameter selection. We therefore used all available 
subjects for training and tune hyperparameters based on convergence assessment and overfitting 
diagnostics, by monitoring ELBO trajectories, stability of the estimated population parameters, and posterior predictive fit.
The average training time was approximately 20 minutes.